\documentclass[10pt,twocolumn]{article}
\usepackage{times}
\usepackage{booktabs} 
\usepackage{amsfonts}
\usepackage{graphicx}
\usepackage{balance}  
\usepackage{color, graphicx, MnSymbol, hyperref} 
\usepackage{balance, multirow, framed, caption, subcaption}  
\usepackage{aliascnt,enumitem}
\usepackage{relsize}
\usepackage{listings}
\usepackage{algpseudocode,algorithm,algorithmicx}
\usepackage{tikz}
\usepackage{textcomp}
\usepackage{pgfplots}
\usepackage{filecontents}
\usepackage[export]{adjustbox}
\usepackage{mathtools}
\usetikzlibrary{arrows,shapes,backgrounds,patterns,decorations.pathreplacing,decorations.pathmorphing,decorations.markings,shadows,shapes.misc,calc,positioning,intersections}

\newcommand{\proto}{{{\tt ProtoXplore}}}
\newcommand{\protoG}{{{\tt Proto-Graph}}}
\newcommand{\admin}{{admin}}
\newcommand{\IE}{{\tt IE}}
\newcommand{\DE}{{\tt DE}}
\newcommand{\BE}{{\tt BE}}
\newcommand{\VC}{{\tt VC}}
\newcommand{\IF}{{\tt IF}}
\newcommand{\DOR}{{\tt DOR}}

\newcommand{\SWT}{{\tt SWT}}

\newcommand{\SMWT}{{\tt SMWT}}
\newcommand{\EMWT}{{\tt EMWT}}
\newcommand{\NWT}{{\tt NWT}}

\newcommand{\RATP}{{\tt RATP}}
\newcommand{\RA}{{\tt RA}}

\newcommand{\FC}{{\tt FC}}

\newcommand{\WT}{{\tt WT}}
\newcommand{\CT}{{\tt CT}}

\newcommand{\tq}{{target query}}
\newcommand{\sq}{{source query}}
\newcommand{\ts}{{target stage}}
\newcommand{\sos}{{source stage}}

\newcommand{\FAIR}{{\tt FAIR}}

\newcommand{\targetstage}{{\tt ts}}
\newcommand{\targetquery}{{\tt tq}}
\newcommand{\sourcestage}{{\tt ss}}
\newcommand{\sourcequery}{{\tt sq}}
\newcommand{\expl}{{\phi}}
\newcommand{\prule}{{\Re}}
\newcommand{\pruletype}{{rule{-}type}}
\newcommand{\pquery}{{query{-}id}}
\newcommand{\pquerytype}{{query{-}type}}
\newcommand{\pvalue}{{rule{-}value}}
\newcommand{\porder}{{rule{-}order}}
\newcommand{\id}{{\tt id}}

\newcommand{\host}{{\tt host}}
\newcommand{\resource}{{\tt res}}
\newcommand{\subresource}{{\tt res'}}
\newcommand{\pathimpact}{{\mathcal{P}}}

\newcommand{\tab}{\hspace*{2em}}

\newcommand{\proj}[1]{{\Pi}}
\newcommand{\sel}[1]{{\sigma}}

\newcommand{\cut}[1]{}
\newcommand{\cutfull}[1]{}

\newcommand{\commentresolved}[1]{}

\newcommand{\ie}{{\it i.e.}} 
\newcommand{\eg}{{\it e.g.}} 

\newcounter{challenge}
\newenvironment{challenge}[1][]{\refstepcounter{challenge}\par\medskip
	\noindent \textbf{Challenge~\thechallenge. #1} \rmfamily}{\medskip}

\newaliascnt{conjecture}{theorem}			
\aliascntresetthe{conjecture}  				
\newaliascnt{remark}{theorem}				
              
\aliascntresetthe{remark}  					
\newaliascnt{corollary}{theorem}			
\aliascntresetthe{corollary}  				
\newaliascnt{definition}{theorem}			
\newtheorem{definition}[definition]{Definition}    
\aliascntresetthe{definition}  				
\newaliascnt{proposition}{theorem}			
\newtheorem{proposition}[proposition]{Proposition}  
\aliascntresetthe{proposition}  				
\newaliascnt{example}{theorem}			
\newtheorem{example}[example]{Example}  	
\aliascntresetthe{example}  				
\newaliascnt{observation}{theorem}			
\aliascntresetthe{observation}  				

\cut{

}

\begin{document}


\title{Analyzing Query Performance and Attributing Blame for Contentions in a Cluster Computing Framework}

\author{Prajakta Kalmegh, Shivnath Babu, Sudeepa Roy \\
	\small {\em  Department of Computer Science, Duke University } \\
		\small {\em  308 Research Drive Duke Box 90129, Durham, NC 27708-0129 } \\
	\small {\em \{pkalmegh,shivnath,sudeepa\}@cs.duke.edu  } \\ [2mm]
}
\date{}
\maketitle

\begin{abstract}
	\vspace{-4mm}
There are many approaches is use today to either prevent or minimize the impact of inter-query interactions on a shared cluster. Despite these measures, performance issues due to concurrent executions of mixed workloads still prevail causing undue waiting times for queries. Analyzing these resource interferences is thus critical in order to answer time sensitive questions like \textit{who is causing my query to slowdown} in a multi-tenant environment. More importantly, dignosing whether the slowdown of a query is a result of resource contentions caused by other queries or some other external factor can help an admin narrow down the many possibilities of performance degradation. This process of investigating the symptoms of resource contentions and attributing blame to concurrent queries is non-trivial and tedious, and involves hours of manually debugging through a cycle of query interactions. \par

In this paper, we present \proto\  - a \textit{Proto} or first system to \textit{eXplore} contentions, that helps administrators determine whether the blame for resource bottlenecks can be attibuted to concurrent queries, and uses a methodology called \textbf{R}esource \textbf{A}cquire \textbf{T}ime \textbf{P}eriod (RATP) to quantify this blame towards contentious sources accurately. Further, \proto\ builds on the theory of explanations and enables a step-wise deep exploration of various levels of performance bottlenecks faced by a query during its execution using a multi-level directed acyclic graph called \protoG. Our experimental evaluation uses \proto\ to analyze the interactions between TPC-DS queries on Apache Spark to show how \proto\ provides explanations that help in diagnosing contention related issues and better managing a changing mixed workload in a shared cluster. \par

\cut{We construct a multi-level directed acyclic graph called \protoG\ to formally capture different types of explanations that link the performance of concurrent queries. In particular, (a) we designate the components of a query's lost (wait) time as \textit{Immediate Explanations} towards its observed performance, (b) represent the rate of contention per machine as \textit{Deep Explanations}, and (c) assign responsibility to concurrent queries through \textit{Blame Explanations}. We develop new metrics to accurately quantify the impact and distribute the blame among concurrent queries. 

Our metrics for impact analysis on \protoG\ enable us to generate rules for a cluster scheduler that can be used to improve the performance of recurring workloads. In our experiments, we use these rules to intervene the task placement strategies for a microbenchmark workload using TPCDS queries on Apache Spark, and show improvement in the performance of victim queries in recurring executions. 

We perform an extensive experimental evaluation using \proto\ to analyze the query interactions of TPC-DS queries on Apache Spark using microbenchmarks illustrating the effectiveness of our approach, and illustrate how the output from \proto\ are used by alternate scheduling and task placement strategies to help improve the performance of affected queries in recurring executions. }
\cut{
Our experiments show that \proto\ performs impact analysis and generates accurate blame attributions efficiently. We also present how the output from \proto\ can be used by alternate scheduling and task placement strategies to help improve the performance of affected queries in recurring executions. 
}
\end{abstract}

\cut{
Analyzing performance interference due to concurrent query executions is challenging on cluster computing frameworks. Under contentions caused by different queries, it is typical to see tasks of the same query executing on the same machines with similar data properties to exhibit varying performance. Identifying symptoms of such erratic behavior, and analyzing the impact of contentions on the overall query execution is tricky as (i) queries are processed as dataflows of multiple stages with several parallel chains of execution, and (ii) performance isolation is hard to achieve on shared clusters even with reservations and constrained capacities. Today no tool exists to help an \admin\ perform a deep analysis of resource contentions. \par

To this effect, we present \proto\ - a \textit{\textbf{Proto}} or first system to \textit{e\textbf{X}plore} the interactions between concurrent queries on a cluster computing framework. We construct a multi-level directed acyclic graph of explanations to formally capture different types of explanations that link the performance of concurrent queries. In particular, (a) we designate the components of a query's lost (wait) time as \textit{Immediate Explanations} towards its observed performance, (b) represent the rate of contention per machine as \textit{Deep Explanations}, and (c) assign responsibility of impact to concurrent queries through \textit{Blame Explanations}. We develop new metrics to quantify the impact and distribute the blame among concurrent queries. We performed an extensive experimental evaluation using \proto\ to analyze the query interactions of TPC-DS queries on Apache Spark using microbenchmarks and user experience studies. Our experiments show that \proto\ performs impact analysis and generates accurate blame attributions efficiently. We also present how the output from \proto\ can be used by alternate scheduling and task placement strategies to help improve the performance of affected queries in recurring executions. 

}

\vspace{-7mm}
\section{Introduction}\label{sec:intro}
\vspace{-4mm}

There is a growing demand in the industry for autonomous data
processing systems \cite{oracle-autonomous-database}. Some critical
roadblocks need to be addressed in order to make progress towards
achieving the goal of automation. One of these roadblocks is in
ensuring predictable query performance in a multi-tenant system. `` \textit{Why
is my SQL query slow?}" This question is nontrivial to answer in many
systems. The authors have seen firsthand how enterprises use an army
of support staff to solve problem tickets filed by end users whose
queries are not performing as they expect. An end user can usually
troubleshoot causes of slow performance that arise from her query
(e.g., when the query did not use the appropriate index, data skew, change in execution plans, etc.). 
However, often the root cause of a poorly-performing `\textit{victim query}' is resource
contention caused by some other `\textit{culprit queries}' in a multi-tenant
system \cite{CPI,BadAWS,BadAmazon,SlowNodes,StolenTime}. Diagnosing such causes
is hard and time consuming. Today, cluster administrators have to manually 
traverse through such intricate cycles of query interactions to identify 
how resource interferences can cause performance bottlenecks. \par

\vspace{1.5mm}
\noindent 
{\bf Should the solution be prevention, cure, or both?}
Features like query isolation at the resource allocation level provide
one approach to address the problem of unpredictable query performance
caused by contention. For example, an \admin\ tries to reduce conflicts by partitioning resources among tenants using capped capacities \cite{CapacitySched}, reserving shares \cite{Rayon} of the cluster, or dynamically regulating offers to queries based on the configured scheduling policies like FAIR \cite{DBLP:journals/pvldb/0002M13} and First-In-First-Out (FIFO). Despite such meticulous measures, providing performance isolation guarantees is still challenging since resources are not governed at a fine-granularity. Moreover, the allocations are primarily based on only a subset of the resources (only CPU in Spark \cite{Spark}, CPU and Memory in Yarn \cite{Yarn}) leaving the requirements for other shared resources unaccounted for. For example, two queries that are promised equal shares of resources get an equal opportunity to launch their tasks in Spark. However, there are no guarantees on the usages of other resources like memory, disk IO, or network bandwidth for competing queries. For instance, if a task uses more resources due to skewed input data on a single host, this can result in higher resource waiting times for tasks of other queries on this host compared to others.  \par

However, in practice, an approach based solely on prevention will not
solve the problem. Real-life query workloads are a mix of very diverse
types of queries. For example, long-running ETL
(Extract-Transform-Load) batch queries often co-exist with short
interactive Business Intelligence (BI) queries. Analytical SQL queries are submitted along with machine learning, graph analytics, and data mining queries. It is hard to forecast
what the resource usage of any query will be and to control what mix
of queries will run at any point of time. In addition, the end users
who create and submit the queries may have very different skill
levels. All end users may not be skilled enough to submit well-tuned
queries. Thus, a practical approach to
provide predictable query performance should supplement `\textit{prevention}'
techniques with techniques for `\textit{diagnosis and cure}'. 

\vspace{1.5mm}
\noindent 
{\bf Our Contributions:} This paper focuses
on the latter. We present \proto, - a Proto or first system to eXplore the resource interferences between concurrent
queries that provides a framework to attribute blame for contentions
systematically. \proto 's foundations are based on the theory of
providing explanations. While there have been several attempts to drill-down to the 
systemic root causes when a query slows down, we believe that \proto\ is a first attempt 
towards using symptoms of low-level resource contentions to analyze high-level inter-query interactions. 
Specifically, we make the following contributions: \textbf{(a) Blame Attribution:~} We use the Blocked Times~\cite{BlockedTime} values for multiple resources (CPU, Network, IO, Memory, Scheduling Queue) to develop a metric called \emph{Resource Acquire Time Penalty} ($\RATP$) that aids us in computing \textit{blame} towards a concurrent task.\textbf{(b) Explanations using \protoG:~} We present a multi-level Directed Acyclic Graph (DAG), called \protoG, that enables consolidation of $\RATP$ and blame values at different granularity. \cut{\textbf{(c) Blame Analysis:~} \protoG\ enables administrators to perform four concrete use-cases in detecting: (i) hot resources, (ii) slow nodes, (iii) high impact \textit{causing} culprit queries, and (iv) high impact \textit{receiving} victim queries. \textbf{(d) Rules for Cluster Scheduler:~} We use the top-$k$ explanations in generating alternative query placement rules that can be consumed by an \admin\ manually to tweak workload schedules or applied automatically (a focus of our on-going research) by a cluster scheduler in an online execution.} \textbf{(c) Web-based UI for deep exploration:~} Our web-based front-end (to be demonstrated in SIGMOD'18 \cite{Proto-demo}) allows users to explore \proto\ and find answers for detecting: (i) hot resources, (ii) slow nodes, (iii) high impact \textit{causing} culprit queries, and (iv) high impact \textit{receiving} victim queries.. \par 

Our experiments evaluate \proto\ using TPCDS queries on Apache Spark. The results of our user-study~\cite{Proto-user-study} shows how \proto\ helps user of different expertise to reduce debugging effort significantly. 

\cut{
\vspace{-1.5mm}
\begin{itemize}[leftmargin=*]
	\itemsep0em
	\item \textbf{Blame Attribution:~} We develop a metric called \emph{Resource Acquire Time Penalty} ($\RATP$) to capture the wait-time distributions of a query for every resource. Using \RATP\ of a task as a basis, we develop a metric to compute the contribution of concurrent tasks in causing contention for this task. \cut{We show how this metric compares with other prevalent measures and helps us avoid false attributions. }
	
	\item \textbf{Explanations using \protoG:~} We present a multi-level Directed Acyclic Graph (DAG), called \protoG, that enables consolidation of $\RATP$ and blame values at different granularity. Using \protoG, we generate three levels of explanations for a query's slowdown due to resource interferences, namely Immediate Explanations (\IE), Deep Explanations (\DE), and Blame Explanations (\BE) that lets users identify sources of contentions at resource, host and query levels respectively. 
	
	\cut{
	
	\begin{itemize}[leftmargin=*]
		\vspace{-3mm}
		\itemsep0em
		\item \textbf{Immediate Explanations:~} identify disproportionate waiting times a query spends for a particular resource,
		\item \textbf{Deep Explanations:~} inspect the wait-time distributions for each resource \cut{used by the query }on specific cluster hosts, 
		\item \textbf{Blame Explanations:~} investigate the contribution of concurrent queries and their stages in causing this wait-time for a particular resource and host combination. 
		\vspace{-3mm}
	\end{itemize}
	Thus, \protoG\ aids a user to perform a deep exploration of resource interferences due to concurrent queries.
}	
	\item \textbf{Blame Analysis:~} \cut{We further define a \emph{Degree of Responsibility} (\DOR) metric to assign blame or responsibility to each node in \protoG\ for causing contention to a query under consideration. }Our interface for \protoG\ enables administrators of cluster computing systems to perform four concrete use-cases in detecting: (i) hot resources, (ii) slow nodes, (iii) high impact \textit{causing} culprit queries, and (iv) high impact \textit{receiving} victim queries. 
	
	\item \textbf{Rules for Cluster Scheduler:~} We use the top-$k$ explanations output at each level by our blame analysis API in generating alternative query placement and query priority readjustment rules. We demonstrate how these rules can be used by an \admin\ to manually tweak workload schedules or applied automatically (a focus of our on-going research) by a cluster scheduler in an online execution. 
	 \cut{We also show how a cluster scheduler can use the heuristics output by our rule generation for tweaking with the workload schedules.}
	
	\item \textbf{Evaluation and User-Study:~} We conduct correctness, efficacy, scalability, and instrumentation overhead analysis experiments to evaluate \proto\ using TPCDS queries on Apache Spark. The results of our user-study involving participation from users of different expertise helps substantiate our claims of accuracy. The study also shows how \proto\ saves users a huge time compared to the alternatives available today.    
	
	\item \textbf{Web-based UI for deep exploration:~} Analyzing contentions step-wise in a multi-level large scale graph is non-trivial. Our web-based front-end (to be demonstrated in SIGMOD'18 \cite{Proto-demo}) allows users to explore \proto\ and find answers for the above use-cases. 
\end{itemize}
}

\vspace{-3mm}
\subsection{Related Work}
\vspace{-1.5mm}
\noindent 
\textbf{Prevention of Interferences:}
Approaches like \cite{CapacitySched,DBLP:journals/pvldb/0002M13} help an \admin\ control resource allocations in a cluster executing mixed workloads. Techniques like resource reservations\cite{Rayon} or dynamic reprovisioning\cite{Morpheus} isolate jobs from the effects of performance variability caused by sharing of resources. Our on-going research focuses on prescriptive mechanisms using online explanations generated from \proto\ to avoid interferences. 

\vspace{1.5mm}
\noindent 
\textbf{Diagnosis and Cure:}
Figure~\ref{table:related_work} presents a summary of how our work compares to the following approaches:

\noindent
\textit{\textbf{(1) Monitoring Tools:}} Cluster health monitoring tools like Ganglia~\cite{Ganglia} and application monitoring tools like Spark UI \cite{SparkUI} and Ambari~\cite{Ambari} provide an interface to inspect performance of queries. \cut{These tools when used together can help observe symptoms of performance degradation and resource bottlenecks in a cluster. }

\begin{table}[h]
	\small
	\centering
	\vspace*{-4mm}
	\caption{Comparison of \proto\ with other approaches}
	\label{table:related_work}
	\vspace*{-3mm}
	\begin{tabular}{|p{2cm}||p{0.7cm}|p{0.8cm}|p{1cm}|p{0.9cm}|p{0.7cm}|} \hline
		(Category) Related Work& OLAP Workloads & Detect Slow-down & Detect Bottlenecks & Blame Attribution & Data-flow Aware  \\ \hline \hline
		(1) Ganglia, Spark UI, Ambari & $\checkmark$ & & $\checkmark$ & & \\ \hline
		(2) Starfish, Dr. Elephant, OtterTune & $\checkmark$ & $\checkmark$ & $\checkmark$ &  & \\ \hline 
		(3) PerfXplain, Blocked Time, PerfAugur & $\checkmark$ & $\checkmark$ & $\checkmark$ & &  \\ \hline
		(3) Oracle ADDM, DIADS& $\checkmark$ & $\checkmark$ & $\checkmark$ & & $\checkmark$  \\ \hline
		(4) CPI$^{2}$ & & $\checkmark$ & $\checkmark ^{CPU}$ & $\checkmark$ &  \\ \hline
		(4) DBSherlock& & $\checkmark$ & $\checkmark$ & & $\checkmark$ \\ \hline \hline
		\tab  \proto\ & $\checkmark$ & $\checkmark$ & $\checkmark$ & $\checkmark$ & $\checkmark$ \\ \hline
	\end{tabular}
	\vspace{-4mm}
\end{table}

\noindent
\textit{\textbf{(2) Configuration recommendation: }}
More recent tools like Starfish\cite{Starfish}, Dr.Elephant\cite{dr_elephant}, and OtterTune\cite{OtterTune} analyze performance of dataflows and suggest changes in configuration parameters. While these approaches helps in identifying and fixing problems in configurations, it is hard to predict how these system-wide changes will affect the resource interferences from inter-query interactions in an online workload. 

\cut{the impact of these system-wide changes on the new patterns of resource interferences in an online workload. that can consequently arise in a shared cluster with continually arriving adhoc or best-effort jobs. Moreover, going beyond investigating the symptoms of contentions and assessing the contribution of concurrently running queries towards causing this impact still remains challenging. }

\noindent
\textit{\textbf{(3) Root Cause Diagnosis tools:}} Performance diagnosis has been studied in the database community \cite{OracleADDM,Borisov,DBSherlock}, for cluster computing frameworks \cite{Perfxplain,BlockedTime}, and in cloud based services \cite{Perfaugur}. The design of \proto\ is based on several concepts introduced in these works: \textbf{(a) Database Community:} In \textbf{ADDM} \cite{OracleADDM}, \textit{Database Time} of a SQL query is used for performing an impact analysis of any resource or activity in the database. \proto\ furthers this approach to provide an end-to-end query contention analysis platform while also enabling multi-step deep exploration of contention symptoms. \proto 's multi-level explanations approach is motivated from DIADS \cite{Borisov}. They use Annotated Plan Graphs that combines the details of database operations and Storage Area Networks (SANs) to provide an integrated performance diagnosis tool. They, however, do not consider any factors related to the impact of these queries in affecting the values of low-level metrics. We believe that this is really important to do an accurate impact analysis. The problem addressed in DIADS is not related, but our multi-level explanation framework bears similarity to their multi-level analysis. Causality based monitoring tools like \textbf{DBSherlock} \cite{DBSherlock} use causal models to perform a root cause analysis. DBSherlock analyzes performance problems in OLTP workloads, whereas \proto\ focuses on performance issues caused by concurrency in data analytical workloads. \textbf{(b) Cluster Computing:} \textbf{PerfXplain} is a debugging toolkit that uses a decision-tree approach to provide explanations for the performance of MapReduce jobs. \proto\  considers dataflow dependencies and workload interactions to generate multi-granularity explanations for a query's wait-time on resources. \textbf{Blocked Time} metric \cite{BlockedTime} emphasizes the need for using resource waiting times for performance analysis of data analytical workloads; we use this pedestal to identify the role of concurrent query executions in causing these blocked times for a task. \cut{We also illustrate with scenarios why we need a more evolved metric compared to \textit{blocked time} for blame attribution.} \textbf{(c) Cloud:} \textbf{PerfAugur} \cite{Perfaugur}  identifies systemic causes for query slowdown. Finding the root cause for slowdown is \textit{not} the focus of \proto. Our motivation is to enable an \admin\ dignose \textit{whether} and \textit{why} this slowdown is a result of resource contentions caused by other queries or some other external factor.

\vspace{-1mm}
\noindent
\textbf{\textit{(4) Blame Attribution: }} \\
CPI$^{2}$ \cite{CPI} uses hardware counters for low-level profiling to capture resource usage by antagonist queries while the CPI (CPU Cycles-Per-Instruction) of the victim query takes a hit. Since this approach does not capture multi-resource contentions at application-level, it suffers from finding poor correlations when queries impact on other resources. Blame attribution has also been studied in the context of \emph{program actions} \cite{BlameAttribution}. The purpose and application addressed in \proto\ is totally different. 

\vspace{-1mm}
\noindent
\textbf{\textit{(5) Other work on explanations in databases:}} The field of explanations has been studied in many different contexts like data provenance \cite{Buneman+2001}, causality and responsibility \cite{MeliouGMS2011}, explaining outliers and unexpected query answers \cite{DBLP:journals/pvldb/0002M13, RoyS14}, etc. The problem studied and the methods applied in this paper are unrelated to these approaches.

\vspace{2mm}	
\noindent
\textbf{Roadmap.~} In Section~\ref{sec:challenges} we describe the key challenges in analyzing dataflows, multi-resource contentions and attributing blame in a shared cluster. We present our blame attribution process in Section~\ref{sec:blame}. We describe the construction of \protoG\ in Section~\ref{sec:framework}. Our experimental and user-study results are in Section~\ref{sec:experiments}, and finally conclude with on-going and future work in Section~\ref{sec:conclusion}.

\vspace{-3mm}

\vspace{-4mm}
\section{Challenges}
\label{sec:challenges}
\vspace{-4mm}

In this section, we review some challenges for blame attribution and discuss the motivation behind our work. \cut{choices of metrics and graph-based model. }

\cut{
\vspace{-1mm}
\subsection{Background: Dataflow Dependencies}
\label{subsubs:data_dependency}
Users of Hadoop \cite{Hadoop} and Spark \cite{Spark} submit applications through high level data processing engines, \eg,  SparkSQL, Dataframes API \cite{SparkSQL}, GraphX \cite{GraphX}, D-Streams \cite{SparkStreams}, MLlib \cite{Mllib}, Hive \cite{Hive}, and Oozie \cite{Oozie}. These applications are processed as a physical execution plan or a \emph{dataflow}, which is a DAG of low-level parallelizable units (\eg, map-reduce \emph{jobs} in Hadoop, \emph{stages} in Spark). In this paper, for simplicity, we adopt the term {\bf stages}. Each edge in the DAG represents the dataflow between these stages\footnote{In practice, an application is decomposed into a DAG of \emph{jobs}; An \textit{action} of a job\cut{, defined by job boundaries,} is executed using a DAG of stages. We do not include this additional layer of jobs in our model since it does not affect our approach and algorithms.}. A stage is composed of a {\textbf{task set}} where each {\textbf{task}} performs the same transformation in parallel on different blocks of the dataset. Each stage starts upon completion of all its parent stages in the dataflow DAG. The root stage of the DAG is the final stage, whereas the leaf stages represent the stages scanning input data. \cut{The {\bf depth} of a stage in the query DAG is counted as the maximum number of stages on a directed path from that stage to the root stage in the DAG.} 
We illustrate these concepts with an example:
}
\begin{figure}[h]
	\vspace{-3mm}
	\centering
	\includegraphics[height=0.5in,keepaspectratio]{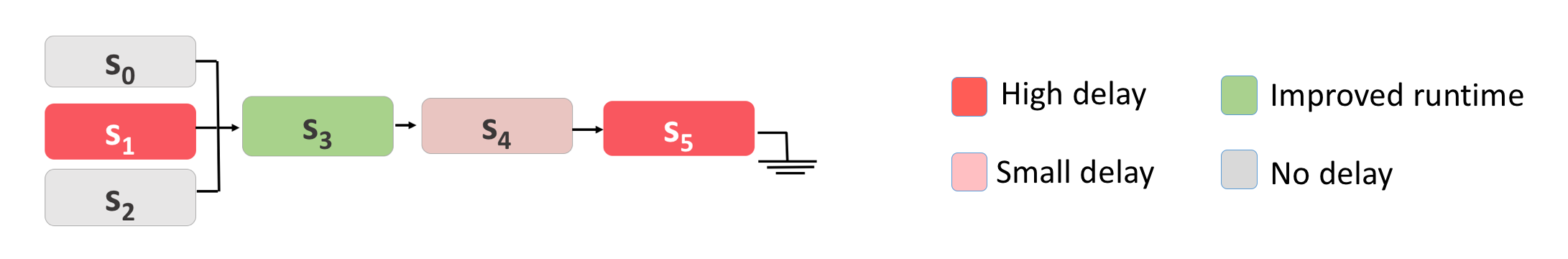}
	\caption{\textbf{\small Dataflow DAG of an example query $Q_0$.
			\cut{Dark red ($s_1, s_5$) denotes a large delay in the expected runtime, light red ($s_4$) denotes a small delay, whereas green ($s_3$) denotes a runtime as expected. \cut{Grey ($s_0, s_2$) shows unaffected stages.}}
	}}
	\label{fig:example_dataflow}
	\vspace{-7mm}
\end{figure}

\noindent
\begin{example}\label{eg:query_dag}
	\vspace{-5mm}
	Consider the 
	dataflow DAG of a query $Q_0$ in Figure~\ref{fig:example_dataflow} comprising stages $s_0, s_1, \cdots, s_5$. The stage $s_5$ is the final or root stage, whereas the leaves $s_0, s_1, s_2$ scan input data. The stage $s_3$ can start only when all of $s_0, s_1, s_2$ are completed. \cut{The depth of stage $s_5$ is 1, and the depth of stage $s_1$ is 4. }Suppose an \admin\ notices slowdown of a recurring query $Q_0$ shown in Figure~\ref{fig:example_dataflow} in an execution compared to a previous one, and wants to analyze the contentions that caused this slowdown.
	The \admin\ can use tools like SparkUI \cite{SparkUI} to detect that tasks of stage $s_{1}$ took much longer than the median task runtime on host (\ie, machine) $X$, and then can use logs from tools like Ganglia \cite{Ganglia} to see that host $X$ had a high memory contention. Similarly she notices that $s_5$ was running on host $Y$ that had a high IO contention.
	Further, the \admin\ sees that 
	stage 
	$r_3$ of another query $Q_1$ was executing concurrently with only stage $s_{5}$ of $Q_0$, while stages $u_5, u_7, u_9$ of query $Q_{2}$ were concurrent with both $s_{1}$
	and $s_{5}$. 
	\cut{compared to $s_0$ and $s_2$, whereas $s_3$ executed as expected (shown in green) and $s_4$ 
	had a small delay (shown in light red).}
	\vspace{-4mm}
\end{example}

\begin{challenge}\label{sub:challenge1}
	\textbf{Assessing Performance bottlenecks:} \newline 
	\textit{Use of blocked time \cite{BlockedTime} alone to compare impact due to resource contentions can be misleading. }
\vspace{-2mm}
\end{challenge}

\noindent
Suppose both the \ts s $s_1$ and $s_5$ spent the same time waiting for a resource (say, CPU); the impact of contention faced by them could still differ if they processed different sizes of input data. Identifying such disproportionality in wait-times can help an \admin\ diagnose victim stages in a dataflow . \cut{In Section~\ref{sec:blame}, we show how \proto\ captures input data size semantics of each stage in appraising the impact values.} 

\cut{In Example~\ref{eg:running_contention}, if we capture that \sos\ $u_{5}$ \textit{steals} the same amount of CPU time from both \ts s $s_{1}$ and $s_{5}$, it does not imply that both $s_{1}$ and $s_{5}$ faced the same amount of contention due to $u_5$. It is possible that $u_{5}$ affected \ts\ $s_{1}$ on more than one resource (memory and other resources) while it caused only CPU contention to \ts\ $s_{5}$. In addition, stolen time does not consider the resource requirements of each task executing on the same machines. The total CPU time that the \sos\ $u_{5}$ has stolen from \ts\ $s_{5}$ will have a higher impact if the amount of input data for $s_{5}$ was less (since it is a root stage) than that of $s_{1}$ (since it is a leaf stage). \par}

\begin{challenge}\label{sub:challenge2}
	\textbf{Capturing Resource Utilizations:} \newline
	\textit{Low-level resource usage trends captured using hardware counters cannot be used towards blame attribution for queries.}
\vspace{-2mm}
\end{challenge}

\noindent
Approaches like capturing the OS level resource utilizations~\cite{CPI} require computing the precise acquisition and release timeline of each resource using hardware counters, and associating them with the high-level abstract task entities. This is non-trivial especially for resources like network (as requests are issued in multiple threads in background) and IO (requests are pipelined by the OS). \cut{Instead, in \proto, we use the wait-time distribution of tasks for each resource for this association. This simpler choice enables us to deduce and attribute fault as we show in Section~\ref{sec:blame}.}

\cut{Two concurrent tasks can compete for shared resources like fetching data from the network buffer queues, reading or writing data to disk, or even multiplexing between CPU cycles. The contention for CPU cycles is very common in executor-based models like Spark where tasks are launched in long-running threads. Tasks also block each other for accessing execution memory (\eg\ sorting
	, etc.) and for storing input data. }

\cut{A typical approach to identify \cut{\textit{source queries} that cause the tasks of a \tq\ to face contentions for these resources}contention causing tasks in shared clusters is to look at the utilization patterns of concurrent tasks~\cite{CPI} for each resource. It is difficult to adopt this course for our purpose as capturing the low-level \textit{resource usage} trends and associating them with the high-level abstract task entities is extremely non-trivial. }

\begin{figure}[h]
	\vspace{-2mm}	
	\centering
	\includegraphics[width=2.8in,keepaspectratio]{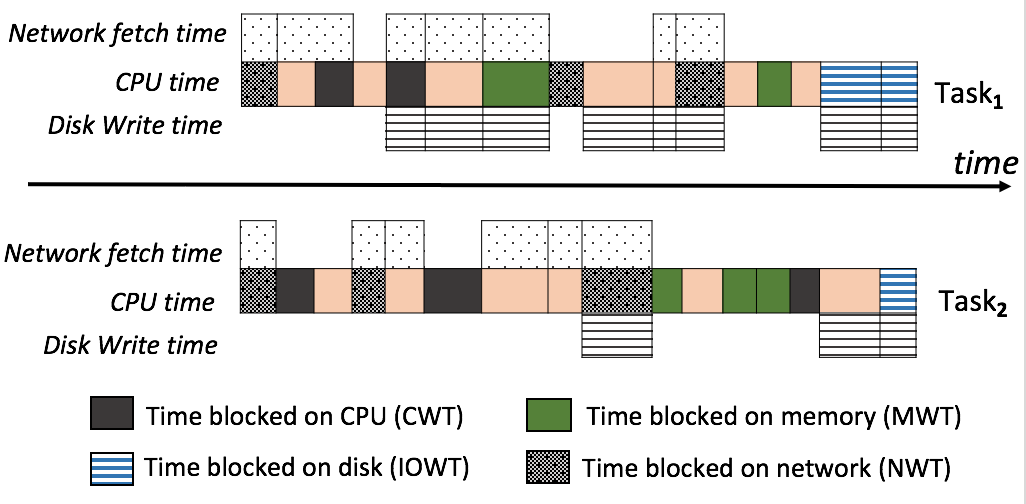}
	\caption{\textbf{Example overlap of resource usage between $Task_{1}$ and $Task_{2}$. Notice the multiplexing between the tasks for compute time.}}
	\label{fig:two_task_overlap}
	\vspace{-5mm}
\end{figure}
\vspace{-1mm}
\begin{challenge}\label{sub:challenge3}
	\textbf{Capturing Resource Interferences:} \newline
	\textit{CPU Time stolen from concurrent queries is inadequate to capture multi-resource contentions.}
\vspace{-1.5mm}
\end{challenge}

\noindent
As shown in Figure ~\ref{fig:two_task_overlap}, tasks can use and be also blocked on multiple resources simultaneously \cite{BlockedTime} resulting in multi-resource interferences between concurrent tasks. As a result, metrics like \textit{stolen time} \cite{StolenTime} cannot be used \textit{as is} to quantify blame as it is used only in the context of identifying processes that steal CPU time. Capturing time \textit{stolen} for multiple resources simultaneously requires a more involved metric for blame attribution. 

 \cut{Suppose stage $s_{1}$ of $Q_0$ was a victim of a multi-resource contention (IO, CPU and memory) since it was the initial stage in the dataflow.while stage $s_5$ is a victim of a single resource contention (say CPU) since it read and processed minimal data over the network. The need to detect queries that steal such multiple resources from $Q_0$ simultaneously requires a more involved metric for blame attribution. }

\cut{Detecting noisy neighbors is a well-defined problem in virtualized environments \cite{BadAWS}. The concept of \textit{stolen time} (time stolen by other processes from the CPU cycles of a victim process) \cite{StolenTime} evolved with the need to quantify contentions on a cluster, but has been used only in the context of identifying processes that steal CPU time.}

\begin{challenge}\label{sub:challenge4}
	\textbf{Avoiding False Attributions:} \newline
	\textit{Overlap Time between tasks of a query does not necessarily signal a contention for resources between them.}
	\vspace{-1.5mm}
\end{challenge}

\noindent
Today, \admin s consider only the \% overlap between concurrent queries to assign blame, which may lead to faulty attributions. For example, even if stage $s_5$ of $Q_{0}$ had total overlap with IO-intensive tasks of $Q_{2}$, this contention may not have impacted $s_5$ owing to its minimal IO activity (later stages in data analytical queries typically process less data). 

\cut{It is possible that during the exact overlap period, the two tasks were using different resources while getting blocked by a third concurrent task on their respective resources. Such wrong attributions should be averted. Consider another scenario: concurrently running tasks on a host may belong to (i) the same stage, (ii) a different stage of the same query, or (iii) a different stage of a different query. Therefore, each task can get blocked and compete for resources at different time during its execution compete either with its own \textit{`fellow tasks'} (case (i) and (ii)) or with \textit{`competing tasks'} (case (iii)). It is thus also important to determine whether the contentions faced by a query are on its own worth or a fault of other to avoid inaccurate accusations. }

\cut{ In Example~\ref{eg:running_contention}, suppose only the \textit{trailing} tasks of stage $s_{5}$ executing on machine $Y$ on equal amounts of input data faced IO contention. Since this resulted in a longer runtime of these tasks, 
it appears that the source queries whose tasks were running concurrently only with the trailing tasks of $s_{5}$ should get high blame. However, if these trailing tasks were reading a much bigger data compared to other tasks, then this attribution is erroneous. Use of $\RATP$ for the \VC\ values at Level-3 (\DE) helps assign a low responsibility towards this contention, thus mitigating the impact coming from its concurrent queries, and prohibiting assignment of incorrect blame to other queries.
(b) 
Suppose task $t_{s}$ of \sos\ $u_{7}$ had almost 100\% overlap with task $t_{c}$ of \targetstage\ $s_{1}$ on machine $X$. Now, suppose $\RATP$  for $t_{s}$ is higher than $\RATP$  for $t_{c}$ for a particular resource $r$ implying that the time $t_{s}$ had to wait to get a unit of resource $r$ is higher than the time $t_{c}$ had to wait to get a unit of the same resource $r$. Assigning any blame to $u_{7}$  for causing contention in such a scenario is unfair even though $t_{s}$ and $t_{c}$ had 100\% concurrency during execution. The only blame $u_{7}$ deserves is because of the overlap of $t_{s}$ and $t_{c}$ (which we capture using $\FC_{h}$), but it should be minimized as $t_{s}$ itself was a victim of resource contention, which we capture by normalizing the $\RATP$ value of \ts\ by the $\RATP$ of \sos\ in Equation~\ref{eq:leve_4_ec}. \par 
}

\begin{challenge}\label{sub:challenge5}
	\textbf{Analyzing Contentions on Dataflows:} \newline 
	\textit{Dataflow dependencies and parallel stage executions make it harder to diagnose the root cause of overall query slowdown.}
	\vspace{-2mm}
\end{challenge}

\noindent
In the above example, stages $s_1$ and $s_5$ of $Q_0$ (in dark red) were a victim of contention, but there is no easy way for the \admin\ to know which of these stage of $Q_0$ was responsible in its overall slowdown. Additionally, diagnosing whether $Q_1$ or $Q_2$ (and which of their stages) is primarily responsible for creating this contention is hard. 

\vspace{1.5mm}
\noindent
\textbf{Motivation:} 
There are several intricate possibilities even in this toy example, whereas there may be long chains of stages running in parallel in a real production environment making this process challenging. Today \admin s have limited means to analyze the impact due to query-interactions apart from looking at individual cluster utilization logs, specific query logs, and manually identifying correlations in both. This forms the motivation of our work. 

\vspace{-3mm}

\cut{
\vspace{3mm}
\noindent
\textbf{\large Design Goals:} \\
There are several intricate possibilities even in this toy example, whereas there may be long chains of stages running in parallel in a real production environment making this process challenging. As a result, in order to meet the above challenges, we need a framework for contention analysis that exhibits the following properties:
	\begin{itemize}
	\itemsep0em
	\item \textbf{Robust:~} It should be able to capture multi-resource interferences, and distinguish between legitimate and false sources of contention. Further, it should account for other slowdown causes of a query. 
	\item \textbf{Thorough:~} It should enable a step-wise exploration of the contention space. The model should facilitate consolidation of blame and responsibility for different combinations of resources, hosts of execution, and concurrent stages and queries. \cut{A user should be able to analyze the contentions for resources, hosts of execution, and concurrent stages and queries that contribute in these per-host and/or per-resource values of contentions. Finally, a user should be able to aggregate these impacts with the ability to filter or rank them dynamically.}
	\item \textbf{Extensible:~} The framework needs to be extensible to plug-in additional explanation causes (\eg\ wait-times due to garbage collection, etc) or break-down existing explanations (\eg\ IO wait time for scanning data vs. IO wait time for writing data). 
\end{itemize}

Traditional statistical approaches \cite{Ahmad,CPI} to analyze query interactions are inadequate for these goals as they fail to reason about intricate conflicts among concurrent queries and do not facilitate systematic exploration of the space. Our metrics for capturing the impact of contentions enable us to meet the goal for robustness, as we describe in the following section; whereas, the use of a graph based model for step-wise analysis provides a comprehensive and modular framework as we show in Section~\ref{sec:framework}.
}

\cut{When some tasks get delayed because of a high demand for a particular resource (\eg, CPU), they hold on to other resources (\eg, memory) as well, thus causing contention for other concurrently running queries on the already acquired resources. This cycle, as results in a complex cause-effect relationships between resource utilization and runtime of concurrent queries. This also often leads to a cascading effect whereby performing an accurate impact analysis becomes challenging. \par}

\vspace{-2mm}
\section{System Overview}
\label{sec:system}
\vspace{-4mm}

\begin{figure}[h]
	\vspace{-3mm}
	\centering
	\includegraphics[height=2.25in,keepaspectratio]{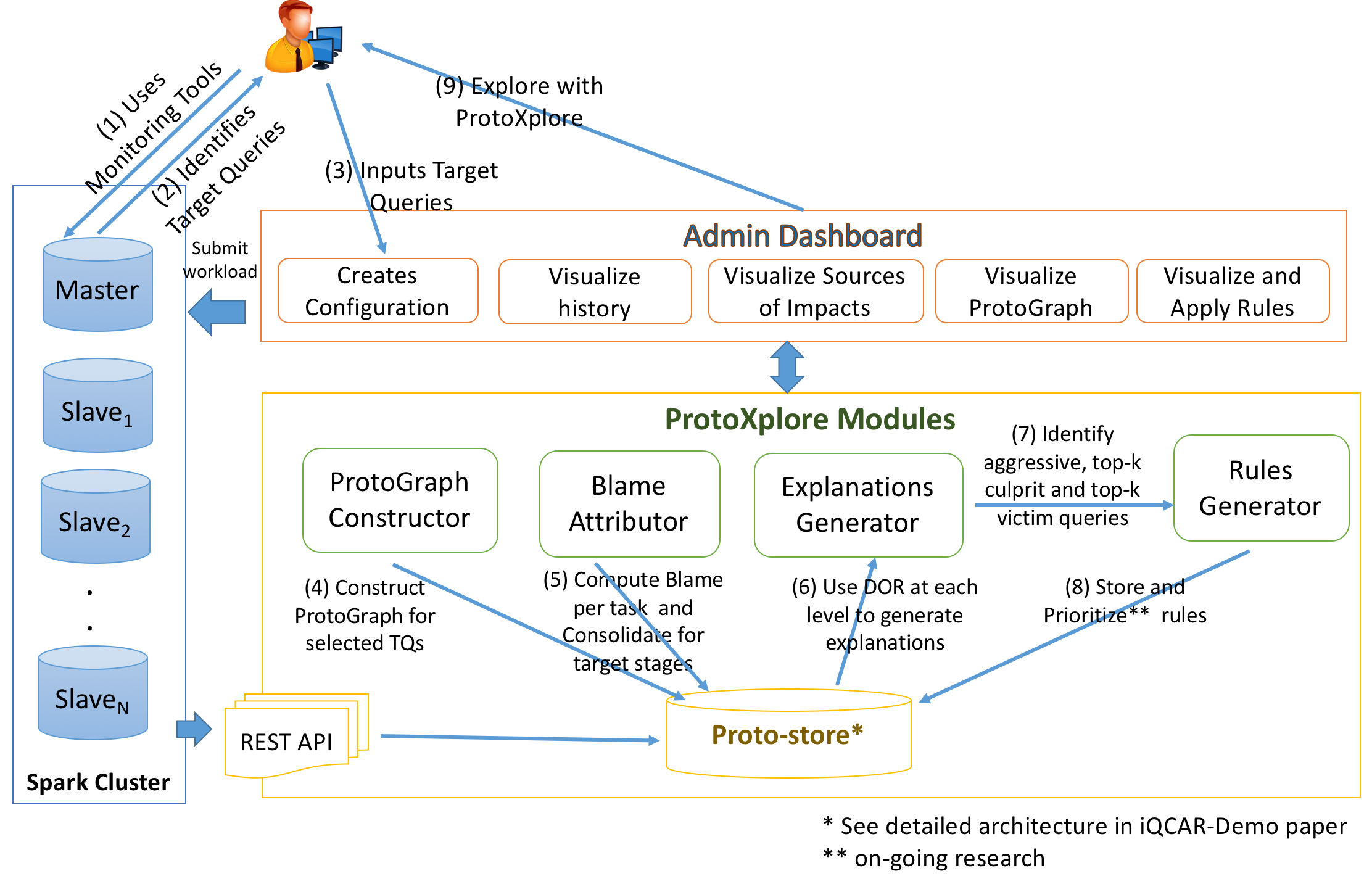}
	\caption{\textbf{\small  \proto\ System Architecture}}
	\label{fig:architecture}
	\vspace{-5mm}
\end{figure}

\proto\ is a system that helps us address the above challenges using our methodology for blame attribution and our graph-based model used for consolidation of blame. It enabes users to detect contentions online while the queries are executing, or perform a deep exploration of contentious scenarios in the cluster offline. Figure~\ref{fig:architecture} shows an overview of \proto 's architecture. 

In Step (1), an \admin\ uses tools like Spark UI~\cite{SparkUI} to monitor the execution of queries and then identifies a set of queries to be analyzed. 
\noindent
\textbf{Target query and target stages:} Any query in the system can be a \textbf{\textit{potential}} source or a target of contention. Each of the queries chosen for deep exploration in \proto\ is termed as a {\bf target query} $Q_{t}$, its stages as {\bf target stages}, and its tasks as {\bf target tasks}. \proto\ currently provides an interface to analyze contentions faced by (i) all stages in the longest path of execution, (ii) any single stage, and (iii) all stages in the \tq (our default option). \cut{To aid the user in this selection, \proto\ accepts a baseline execution as input to recommend queries that took a maximum runtime hit compared to this baseline. Our ongoing research focuses on using query resource usage characteristics to recommend victim queries. For selection of \ts s at Level-1,}

\noindent
\textbf{Source query and source stages:} 
We refer to other concurrently running queries $Q_s$ that can possibly cause contention to a \tq\ $Q_t$ as \textbf{source queries} (if the stage(s) of $Q_s$ are waiting in the scheduler queue or running concurrently with the stage(s) of $Q_t$). Stages of $Q_{s}$ are {\bf \sos}, and its tasks are {\bf source tasks}. Finally, we refer to source queries that cause multi-resource contentions for several concurrent queries with high impact as \textbf{aggressive queries}.

Once the user submits target queries to \proto\ in step (3), the \textbf{\protoG\ Graph Constructor}  builds a multi-level graph as described in Section~\ref{sec:framework} for these queries (Step (4)). In addition, users can configure the resources or hosts for which they want to analyze the contentions. For example, users can diagnose impact of concurrency on only scheduler queue contentions, or CPU contention on a subset of hosts, or source queries submitted by a particular user etc. For each task of a \ts, \proto\ computes the \RATP\ and \textit{blame} metrics as described in Section~\ref{sec:blame}. The \textbf{Blame Attributor} module then consolidates these values for stages, as shown in Step (5), which are then used to update the node and edge weights of the graph subsequently. The \textbf{Explanations Generator} updates a \textit{degree of responsibility} (\DOR)  metric for each node, which is then used by the Blame Analysis API to generate explanations at various granularities (step (6)). Specifically, we generate three levels of explanations for a query's slowdown due to resource interferences, namely Immediate Explanations (\IE), Deep Explanations (\DE), and Blame Explanations (\BE) that lets users identify sources of contentions at resource, host and query levels respectively. These explanations are then consumed in Step (7) by the \textbf{Rule Generator} to output heuristics for the \admin\ (like alternate query placement, dynamic priority readjustment for stages and queries). The details of rule generation is out of scope of this paper and is focus of our ongoing research; a preliminary implementation with basic rules can be found in the demo paper~\cite{Proto-demo}. In step (8), these rules are stored for the admin to later review and apply to the cluster scheduler. The web-based interface presents a dashboard to the admin to visualize cluster and query performance statistics, contention summary graphs for a selected workload, hints in selection of target queries for analysis, top-k sources of impact to the selected target queries, and the corresponding \protoG ~\cite{Proto-demo}. Next, we discuss the theory behind our \textbf{Blame Attributor} module.

\vspace{-6mm}
\section{Blame Attribution}
\label{sec:blame}
\vspace{-3mm}
In order to address the challenges 1-4 described in Section~\ref{sec:challenges}, we develop a metric called Resource Acquire Time Penalty (\RATP) that helps us formulate the blame towards concurrent tasks.

\vspace{-3.5mm}
\subsection{Resource Acquire Time Penalty (RATP)}\label{sec:def-RATP}
In a $\delta \tau$ interval, let the time spent by a target task $tt$ to acquire $\RA_{tt,r,h}^{\delta \tau}$ units of resource $r$ on host $h$ be:
\vspace{-2.5mm}
{\small
\begin{equation}\label{eq:timespent}
T_{tt,r,h}^{\delta \tau}  = \WT_{tt,r,h}^{\delta \tau} + \CT_{tt,r,h}^{\delta \tau}
\end{equation}
}
\vspace{-1.5mm}
where, $\WT_{tt,r,h}^{\delta \tau}$ is the time spent blocked or waiting for the resource and $\CT_{tt}^{\delta \tau}$ is the time spent consuming it in this interval. For the remaining section, we assume that the resource $r$ and host $h$ are fixed, so we omit $r, h$ in the subscripts and other places for simplicity where it is clear from the context. 
\vspace{-1mm}
\begin{definition} \label{def:ratp}
We define the \textbf{Resource Acquire Time Penalty} ({\bf RATP}) for $tt$ for resource $r$ on host $h$\cut{(we omit the $r,h$ subscripts for ease of expression)} in interval $\delta \tau$ as: 
\vspace{-3mm}
{\small
\begin{equation}
RATP_{tt}^{\delta \tau} = \frac{\left( \WT_{tt}^{\delta \tau} + \CT_{tt}^{\delta \tau} \right)}{\RA_{tt}^{\delta \tau}}\label{eq:ratp}
\end{equation}
}
\end{definition}
\vspace{-3mm}

\cut{In Section~\ref{sec:impl}, we elaborate on how we compute this value for different resources used by a task during its execution, namely CPU, IO, Memory, Network and time spent waiting in the scheduling queue.}

\cut{
Let $\tau_{tt}^{start}$ and $\tau_{tt}^{end}$ be the start and end times of task $tt$.
During its runtime $T_{tt} = \tau_{tt}^{end} - \tau_{tt}^{start}$, if we assume a uniform distribution on the wait-time per unit data processed, we get,
{\small 
\begin{equation} \label{eq:ratp_uniform}
	\RATP_{tt, uniform}^{r, h} = \frac{\WT_{tt}^{r,h}}{\RA_{tt}^{r,h}}
\end{equation}
\textit{
 where, $\WT_{tt}^{r,h}$ is the total wait-time of task $tt$ for resource $r$ on host $h$, and $\RA_{tt}^{r,h}$ is the total data processed by $tt$ on this host. }
}

\textbf{\RATP\ Approximation:}
\RATP\ being a function over time can incur high instrumentation cost based on the frequency of measurements. In our implementation, we use an approximate function that is piece-wise uniform based on the following assumption: in a small interval $\delta t$, tasks concurrent to $tt$ continue to cause contention at the same rate unless a qualifying event changes the state of the system.

\begin{sloppypar}
\cut{As  discussed in Table~\ref{table:explanation_relationships} in the appendix,}
\noindent
The type of data processed for different resources are different.
 For example, consider the metric for remote bytes read over the network in Spark ({\scriptsize {\tt REMOTE\_BYTES\_READ}}). The corresponding $\RATP$ metric, called the \emph{`network bytes read penalty'}, is $\frac{\NWT}{{\tt REMOTE\_BYTES\_READ}}$ for a task (omitting the superscripts and subscripts),
and gives us the wait time for processing one unit (one byte for our analysis) of remote data.
\end{sloppypar}
}
 \cut{As another example, $\RATP$ for a task while it waits in the scheduler queue is computed as the time spent waiting (\ie\ \SWT) per number of slots offered to it in this timeframe (\ie\ $\frac{\SWT}{{\tt SLOTS\_OFFERED}}$).}

\cut{
\vspace{3mm}
\noindent
\textbf{Features of RATP:} \label{subsec:ratp_features}
\cut{\RATP\ exhibits the following characteristics addressing the challenges described in Section~\ref{sec:challenges}. }
(1) \textbf{Sensitive:~} Since it captures the wait-time per unit of data processed, it is efficient in detecting disproportionalities\cut{in wait time} discussed in Challenge~\ref{sub:challenge1}.  
(2) \textbf{Computable:~} \cut{Accounting for wait-time of a task based on use times of other tasks is non-trivial as we saw in Challenge~\ref{sub:challenge2}. }\RATP\ relies on availability of wait-times of tasks which are easier to capture at application-level~\cite{BlockedTime} instead of hardware-level instrumentation (see Challenge~\ref{sub:challenge2}). 
(3) \textbf{Resilient:~} \cut{Being a distribution over time} Depending on the frequency at which the wait-time and data usage information is collected, it can be tailored to represent complex resource interference patterns seen in Challenge~\ref{sub:challenge3}. \cut{For instance, historical execution data can be used for recurring workloads to capture interference using \RATP\ with more accuracy. }
(4) \textbf{Comparable:~} The \RATP\ value of two concurrent tasks can be compared using the area under the respective distribution curves for their overlapping period. We show next how we utilize this feature to compute blame.  
}
\vspace{-3.5mm}
\subsection{Slowdown of a task}\label{sec:slowdown}
\vspace{-1mm}
In the $\delta \tau$ interval, let the capacity of a host $h$ to serve a resource $r$ be $\mathcal{C}$ units/sec (\emph{note: $r,h$ superscripts omitted}). Thus, the total used capacity by all consumers of resource $r$ is bounded by the system capacity $\mathcal{C}$, expressed as:
\vspace{-2mm}
{\small
	\begin{equation}\label{eq:capacity}
	\mathcal{C} = \mathcal{C}_{tt} + \mathcal{C}_{1} +\mathcal{C}_{2} + \dots + \mathcal{C}_{n} + \sum_{1}^{k} \mathcal{C}_{ki} + \sum_{1}^{l} \mathcal{C}_{ui}
	\vspace{-1mm}
	\end{equation}
}
where, $\mathcal{C}_{tt}$ is the capacity used by target task $tt$; $\mathcal{C}_{1}, \mathcal{C}_{2}, \dots\ \mathcal{C}_{n} $ are the individual used capacities by $n$ other concurrent tasks; whereas $\mathcal{C}_{ki}$ and $\mathcal{C}_{ui}$ represents capacities used by other $k$ known and $l$ unknown causes. The minimum time required to acquire resource $r$ on this host can then be expressed as $\RATP^{\delta \tau}=\frac{1}{C}$ sec/unit. Thus, slowdown of task $tt$ due to unavailability of resources is:

\begin{definition} \label{def:slowdown}
	Slowdown $\mathcal{S}_{tt}$ of a task $tt$ in interval $\delta \tau$ between time $\tau$ and $\tau + \delta \tau$ is defined as
	\vspace{-2mm}
	{\small
		\begin{equation}\label{eq:slowdown}
		     \mathcal{S}_{tt}^{\delta \tau} = \frac{(\RATP_{tt}^{\delta \tau} - \RATP^{\delta \tau})}{\RATP^{\delta \tau}}
		\end{equation}
\vspace{-1.5mm}
where, $\RATP_{tt}^{\delta \tau}$ is computed as per Equation ~\ref{eq:ratp}.
}
\end{definition}
Intuitively, it is the deviation from the ideal resource acquire time on that host.\cut{The slowdown of $tt$ will be zero when all the resources are available to the task $tt$.} Therefore, the slowdown corresponds to the blame that can be attributed to unavailability of resources which can be attributed to other tasks running concurrently with $tt$ on $h$ during its execution, or on other quantities as shown in Equation~\ref{eq:capacity}, giving us:
 \vspace{-1mm}
	{\small
		\begin{equation}\label{eq:blame_breakdown}
		\mathcal{S}_{tt}^{\delta \tau}= \underbrace{(\sum_{1}^{n} \beta_{st \rightarrow tt}^{\delta \tau})}_{p_1} + \underbrace{\sum_{1}^{k} \beta_{ki \rightarrow tt}^{\delta \tau} + \sum_{1}^{l} \beta_{ui \rightarrow tt}^{\delta \tau}}_{p_2}
	\vspace{-5.5mm}
\end{equation}
}

\noindent		
Here $\beta_{ss \rightarrow ts}^{\delta \tau}$ is the blame assigned to each of the $n$ concurrent tasks executing on the same host during task $tt$'s execution. Wait-times for resources are sometimes also due to processes that are part of the framework but not a valid symptom of  interferences between tasks. (\eg, garbage collection for CPU, HDFS replication for network). These processes impact all concurrent tasks alike for the specific resource. $\beta_{ki \rightarrow tt}^{\delta \tau}$ gives the blame value assigned to other known non-concurrency related causes that result in task $tt$'s wait-time. Finally, slowdown could be due to a variety of other causes which are either not known or cannot be attributed to any concurrent tasks.\cut{For example, as task's \NWT\ is not necessarily more because of a concurrent task's network activity. It may be due to network speed issues, the host from which the data is fetched running slow, etc. } $\beta_{ui \rightarrow tt}^{\delta \tau}$ captures this value of slowdown due to systemic issues. \par

\vspace{-3.5mm}
\subsection{Slowdown due to Concurrency }\label{sec:slowdown_conc}
The slowdown defined in $p_2$ of Equation~\ref{eq:blame_breakdown} is the \textit{best case} slowdown for using resource $r$ on host $h$ in the absence of any interferences due to concurrency. We now derive $p_1$ of Equation~\ref{eq:blame_breakdown}.\cut{, \ie\ blame $\beta_{ss \rightarrow tt}^{\delta \tau}$ from a source stage to a target stage.} First we discuss a simpler case, when there is a full overlap of $tt$ with concurrently running tasks to present the main ideas. Then we discuss the general case with arbitrary overlap between $st$ and $tt$. \par

\vspace{-3.5mm}
\subsubsection{Full overlap of $tt$ with concurrent tasks}
Rewriting Equation~(\ref{eq:capacity}) using $\RATP$ values to account for capacities used by running tasks (omitting the known and unknown causes) in interval $\delta \tau$:

	\begin{subequations}\label{eq:ratp_to_blame}
		\setlength{\abovedisplayskip}{0pt}
		\setlength{\belowdisplayskip}{0pt}
		\begin{align*}
		&\frac{1}{\RATP^{\delta \tau}} = \frac{1}{\RATP_{tt}^{\delta \tau}} + \frac{1}{\RATP_{1}^{\delta \tau}} +	\frac{1}{\RATP_{2}^{\delta \tau}} + \dots + \frac{1}{\RATP_{n}^{\delta \tau}} \\
		&\parbox{22em}{Multiplying by $\RATP_{tt}^{\delta \tau}$ and subtracting 1  on both sides yields,}\\
		&\frac{\RATP_{tt}^{\delta \tau} - \RATP^{\delta \tau}}{\RATP^{\delta \tau}} = 	\frac{\RATP_{tt}^{\delta \tau}}{\RATP_{1}^{\delta \tau}} + \dots + \frac{\RATP_{tt}^{\delta \tau}}{\RATP_{n}^{\delta \tau}} \\ 
		&\parbox{22em}{The left hand side above is the increase in {\RATP} of target task and thus represents its slowdown  $\mathcal{S}_{tt}^{\tau}$ given by (Definition~\ref{def:slowdown})} \\
		&\mathcal{S}_{tt}^{\delta \tau} = \left[ \sum_{st \in\ n} \frac{\RATP_{tt}^{\delta \tau}}{\RATP_{st}^{\delta \tau}} \right]  \tag{\ref{eq:ratp_to_blame}}	
		\end{align*}
	\end{subequations}

Each term on the right hand side of the above equation is contributed by one of the source tasks concurrent to task $tt$,
and corresponds to blame attributable to a source task $st$ in this interval, that is, $\beta_{st \rightarrow tt}^{\tau} = \frac{\RATP_{tt}^{\tau}}{\RATP_{st}^{\tau}}$. 

\vspace{-4.5mm}
\subsubsection{Partial overlap with concurrent tasks}
\begin{figure}[h]
	\vspace{-6mm}
	\centering
	\includegraphics[height=1.5in,keepaspectratio]{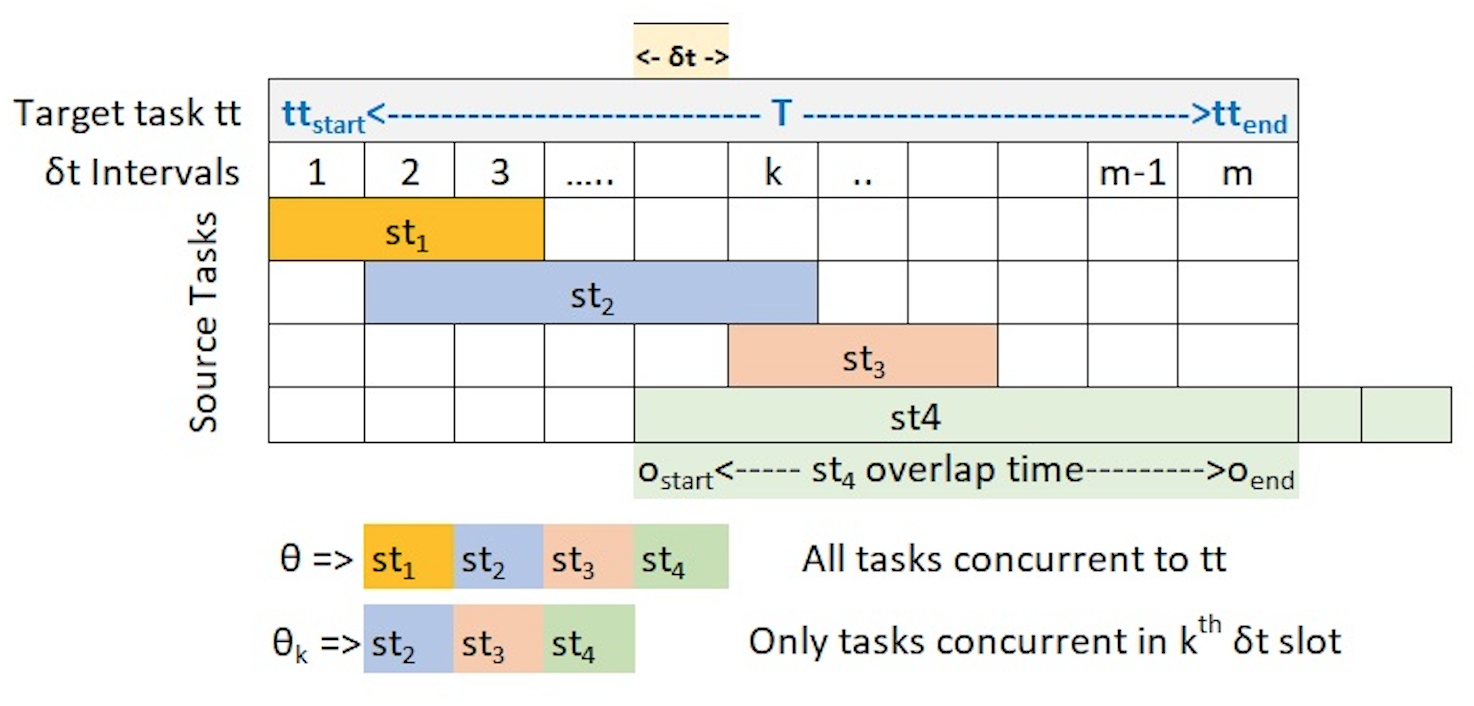}
	\caption{\textbf{\small An example overlap between concurrent tasks.}}
	\label{fig:be_theory}
	\vspace{-5mm}
\end{figure}

\vspace{-2.5mm}
The above derivation works for a time interval in which all concurrent tasks have a total overlap with $tt$. In practice, they overlap for different length of intervals as can be seen from Figure~\ref{fig:be_theory}. As a result, we cannot use Equation~(\ref{eq:ratp_to_blame})  directly to attribute blame to a source task $st$. To derive blame for this scenario, we divide the total duration $T=tt_{end} - tt_{start}$ of task $tt$'s execution time in $\delta \tau$ intervals such that in each $\delta \tau$ time-frame the above  Equation~(\ref{eq:ratp_to_blame}) holds. \par

Let $\mathcal{S}_{1}, \mathcal{S}_{2}, \dots\ , \mathcal{S}_{m}$ be the slowdown in each of the $m=\frac{T}{\delta \tau}$ intervals of the task's execution. This gives us the mean slowdown of $tt$ as ($T_{tt}$ is the execution time of $tt$): 

	\begin{subequations}\label{eq:slowdown_mean_deriv}
		\setlength{\abovedisplayskip}{0pt}
		\setlength{\belowdisplayskip}{0pt}
		\begin{align*}
		&\mathcal{S}_{mean}= \frac{\delta \tau}{T_{tt}} \sum_{k \in m}^{} \mathcal{S}_{k} \\
		&\parbox{23em}{Substituing the value of slowdown $\mathcal{S}_{k}$ in the $k_{th}$ interval using Equation~(\ref{eq:ratp_to_blame}),} \\
		&\mathcal{S}_{mean} = \frac{\delta \tau}{T_{tt}} \sum_{k \in m}^{}  \sum_{st \in \theta_{k}}^{} \frac{\RATP_{tt}^{k}}{\RATP_{st}^{k}} \\
		&\parbox{23em}{where, $\theta_{k} $ is the set of source tasks that are concurrent in the $k_{th}$ interval with task $tt$.}   \\ 
		\vspace{3mm}	
		&\parbox{23em}{In order to understand the contribution of each source task towards the mean slowdown, the double summation can be rearranged as. } \\
		\vspace{3mm}	
		&\mathcal{S}_{mean} = \sum_{st \in \theta}^{}  \sum_{k \in m'}^{} \frac{\RATP_{tt}^{k}}{\RATP_{st}^{k}} \frac{\delta \tau}{T_{tt}} \\
		&\parbox{23em}{where, for each $st$, $m'$ is the number of intervals from time $tt_{start}$ to $tt_{end}$ in which $st$ overlaps with $tt$. The outer sum is now on the set of all tasks, $\theta$, that had an overlap with $tt$. } \\
		\end{align*}
	\end{subequations}

\noindent
The above equation in the limiting case is:
\begin{equation}\label{eq:slowdown_mean}
\mathcal{S}_{mean} = \sum_{st \in \theta}^{} \left[ \int_{o^{start}_{st, tt}}^{o^{end}_{st, tt}} \frac{\RATP_{tt}}{\RATP_{st}} \frac{dt}{T_{tt}}  \right]  
\end{equation}

where,  $o^{end}_{st, tt}$ - $o^{start}_{st, tt}$ is the overlap time between tasks ${tt}$ and ${st}$. The integral inside the summation is the total blame towards a source task $st$, giving us the definition of blame:  \\ 

\vspace{-2mm}
\begin{proposition}\label{prop:blame}
The \textbf{blame} $\beta_{ss \rightarrow tt}^{\delta \tau}$ for the contention caused by a task $st$ of a \sq\ to the task $tt$ of a \tq\ on host $h$ for resource $r$ in interval $\tau$ to $\tau + \delta \tau$ can be expressed as
	{\small
		\begin{equation}\label{eq:blame_attr}
		\beta_{st \rightarrow tt}^{\delta \tau} = \frac{1}{T_{tt}} \left[  \int_{o_{st, tt}^{start}}^{o_{st, tt}^{end}} \frac{\RATP_{tt}^{\delta \tau}}{\RATP_{st}^{\delta \tau}} d \delta \tau \right] 
		\end{equation}
	}
where, $o_{st, tt}^{end}, o_{st, tt}^{start}$ denote the ending and starting points of  the overlap time between tasks $tt$ and $st$ (on host $h$), and 
$T_{tt}$ is the total execution time of task $tt$\cut{ (since a task $tt$ runs on a single host, the superscripts are omitted)}. 
\end{proposition}

\cut{
	\textbf{(1) Assigns just penalties :~} Since the blame is computed only for periods where $tt$ is waiting for resource $r$ during the overlap period, the blame on $st$ is zero if $tt$'s wait-time is zero during the overlap. Moreover, if both $st$ and $tt$ are not waiting for the same resource in the overlapping time (see Challenge~\ref{sub:challenge4}), $st$ does not qualify for blame. Blame attribution using mere overlap time can miss this subtlety.
 \textbf{(2) Accounts for known causes:~} Contentions are sometimes due to processes that are part of the framework but not a valid symptom of  interferences between tasks. (\eg, garbage collection for CPU, HDFS replication for network). These processes impact all concurrent tasks alike for the specific resource. \cut{Their usage can be captured via additional instrumentation or using external tools.} Equation~\ref{eq:blame_breakdown} allows incorporating their impact in the $\beta_{known \rightarrow tt}$ section.
\textbf{(3) Accounts for unknown causes:~} Finally, slowdown could be due to a variety of other causes which are either not known or cannot be attributed to any concurrent tasks. \cut{For example, as task's \NWT\ is not necessarily more because of a concurrent task's network activity. It may be due to network speed issues, the host from which the data is fetched running slow, etc. }Equation~\ref{eq:blame_breakdown}, accommodates all these causes into the $\beta_{unknown \rightarrow tt}$ value. \cut{It can be calculated by subtracting the sum of first two elements from the overall slowdown $\mathcal{S}_{tt}$. }
}

\vspace{-7mm}
\section{Implementation}
\label{sec:impl}
\vspace{-2.5mm}
In this section, we discuss how we derive \textit{blame} in practice, the supported resources for analyzing blame, and the metrics support we have instrumented in Spark. 

\vspace{-3.5mm}
\subsection{Blame Formulation using Blocked Time}
A task continues to execute even when its resource request is not completely fulfilled (especially for IO and Network)~\cite{BlockedTime}. This is because these resources are pipelined to it by background application threads or by the OS. Hence a source task's real impact (attributable blame) is only to the extent that the target task is blocked for the resource. This adjustment is also important to predict the expected speed up of target task more accurately. If an admin decides to kill a source task because of its contention then he can only expect an improvement proportional to target task's blocked time (not its entire resource acquision time). The blocked time $\WT_{tt,r,h}^{\delta \tau}$ values (see Equation~\ref{eq:timespent}) are also relatively more easier to capture as shown in ~\cite{BlockedTime}. We thus define an \textit{adjusted} \RATP\ for a task $tt$ as:

\vspace{-3mm}
{\small
\begin{equation}
\RATP\textit{-blocked }_{tt}^{\delta \tau} = \frac{\WT_{tt}^{\delta \tau} }{\RA_{tt}^{\delta \tau}}\label{eq:ratp_blocked}
\end{equation}
}
For the source task though, we continue to use the entire interval $\delta \tau$ as time spent to acquire $\RA_{st}^{\delta \tau}$ units of resource instead of its blocked time. This lowers the attributable blame further but allows us to account for those sources whose blocked time may not be available (especially terms of $p_2$ in Equation ~\ref{eq:blame_breakdown}) For clarity, we represent this as \textit{max} \RATP\ which for a source $st$ is as:
\vspace{-5mm}

{\small
\begin{equation}
\RATP\textit{-max }_{st}^{\delta \tau} = \frac{\delta \tau}{\RA_{st}^{\delta \tau}}\label{eq:ratp_max}
\end{equation}
}
\noindent
\vspace{1mm}
The blame value we use is therefore:
{\small
	\begin{equation}\label{eq:blame_attr_impl}
	\beta\textit{-blocked }_{st \rightarrow tt}^{\delta \tau} = \frac{1}{T_{tt}} \left[  \int_{o_{st, tt}^{start}}^{o_{st, tt}^{end}} \frac{\RATP\textit{-blocked }_{tt}^{\delta \tau}}{\RATP\textit{-max }_{st}^{\delta \tau}} d \delta \tau \right] 
	\end{equation}
}

As blocked time in an interval is bound by the total interval length, $\beta\textit{-blocked }_{st \rightarrow tt}^{\delta \tau} \le \beta_{st \rightarrow tt}^{\delta \tau}$.  

\vspace{-4mm}
\subsection{Handling Known Causes:} 
\vspace{-2.5mm}
A query may experience slowdown due to myriad reasons like systemic issues (executors getting killed, external processes, etc), the query's own code (used the wrong index, change in query plan, etc.) and many more. The wait-time for every resource can be attributed to any of these issues besides concurrency. On the other hand, concurrenct executions may have additional indirect consequences like slowdown of a task due to increase in garbage collection time owing to a previously executed resource-intensive task, etc. Research in the area of diagnosing such known and unknown systemic causes for performance degradation is well established~\cite{Perfxplain,DBSherlock,Dryad}, which is out of the scope of discussion of this paper. Although \proto 's primary focus is to account for the $\beta_{st \rightarrow tt}^{\delta \tau}$ term, it also comes pre-configured for attributing blame to select known causes like JVMGCTime (cpu), HDFS repartitioning (IO) and RDD storage (Memory) etc. \proto\ also allows adding known causes via its feedback module where users with domain knowledge and prior experience with \proto\ can configure support for them. \proto\ also avoids false attribution by using catch-all term for unknown causes as we show next.\par

\vspace{-4mm}
\subsection{Avoiding False Attributions:}
\label{subsec:blame_features}
\vspace{-2.5mm}
Our quantification of blame $\beta\textit{-blocked }_{st \rightarrow tt}$ enables us to distinguish between legitimate and false sources of contention owing to the following characteristics: (i) Since the blame is computed only for periods where $tt$ is blocked for resource $r$ during the overlap period, the blame on $st$ is zero if $tt$'s wait-time is zero during the overlap. (ii) If both $st$ and $tt$ are not waiting for the same resource in the overlapping time (see Challenge~\ref{sub:challenge4}), $st$ does not qualify for blame. Blame attribution using mere overlap time can miss this subtlety. (iii) Finally, slowdown could be due to a variety of other causes which are either not known or cannot be attributed to any concurrent tasks. To handle these cases, \proto\ creates a synthetic \textit{unknown} source query in the framework. We assume that such a query overlaps with all tasks of every query on every host of execution. For every resource, we then keep a track of the system-level resource used values (currently \proto\ is configured for IO bytes read/written and network bytes read resources) during each execution window. We compute the \textit{total-unaccountable-resource} by subtracting the resource acquired values for all tasks running during that window from the system-level metric. The blame attributed to this \textit{unknown} source is then computed by using the $\RA_{st}^{\delta \tau} =  \textit{total-unaccountable-resource}$ in the $\RATP\textit{-max }_{st}^{\delta \tau}$ computation of Equation~\ref{eq:blame_attr_impl}. \par

\vspace{-4.5mm}
\subsection{Supported Resources}
\label{subsec:resources}
\vspace{-2.5mm}
The values of resource acquired $\RA_{tt}^{\delta \tau}$ for most resources like input or output bytes, shuffle bytes read, etc., and some wait-time metrics like \textit{fetch-wait-time}, \textit{shuffle-write-time} and \textit{scan-time} for a task are available through Spark's default metrics. For additional metrics (denoted with a superscript $^{*}$ against them) and time-series data we added instrumentation to Spark and built per-host agents (for system metrics). In each window of data collected during task $tt$'s execution, we consider contention for the following resources on host $h$: 
\vspace{-2mm}
\begin{itemize}[leftmargin=*]
	\itemsep0em
	\item Network: We compute $\RATP_{network}$ using the Network Wait Time (\NWT) or \textit{fetch-wait-time} metric from Spark. The value of resource acquired $\RA_{network}$ is the shuffle bytes read metric in that window. 
	\item Memory: To capture the contention for \textit{process-managed} memory , we compute (i) $\RATP_{storage-mem}$ using \textit{storage memory wait time} ($\SMWT^{*}$) and \textit{storage memory acquired}$^{*}$, (ii) $\RATP_{exec-mem}$ using \textit{execution memory wait time} ($\EMWT^{*}$) and \textit{execution memory acquired}$^{*}$ metrics. 
	\item IO: We compute the \RATP\ for IO Read $\RATP_{io-read}$ using \textit{scan-time} and bytes read in the scan operation. For IO Write, its $\RATP_{io-write}$ is given using \textit{shuffle-write-time} and \textit{shuffle-bytes-written} metrics in Spark. 
	\item CPU: Apart from above resources, a task's execution can wait due to other reasons like os scheduling wait ,acquiring locks/monitors, priority lowering etc. We capture (a) monitor and lock wait time metrics from JMX (b) garbage collection wait time from spark existing metrics and (c) attribute all other wait time to os scheduling delay. We deal with garbage collection specially by creating a synthetic \textbf{GC} query in our model as a possible source of contention. This query is modeled to have a single long running task on each executor. This task is assumed to block all other tasks during the duration of its execution. We also attribute the remaining unaccounted wait time in an interval to the OS scheduling wait time. As the blame due to a concurrent task is based on $\RATP$ ratios, any error in this way of accounting for scheduling wait is minimized to a large extent. Finally for each of the above CPU-impacting wait-times, we use the value of \textit{executor-cpu-time} as the resource acquired.
\end{itemize}

\vspace{-5.5mm}
\subsection{Frequency of Metrics Collection:} 
\label{subsec:freq}
\begin{figure}[h]
	\vspace{-3.5mm}
	\centering
	\includegraphics[height=1.2in,keepaspectratio]{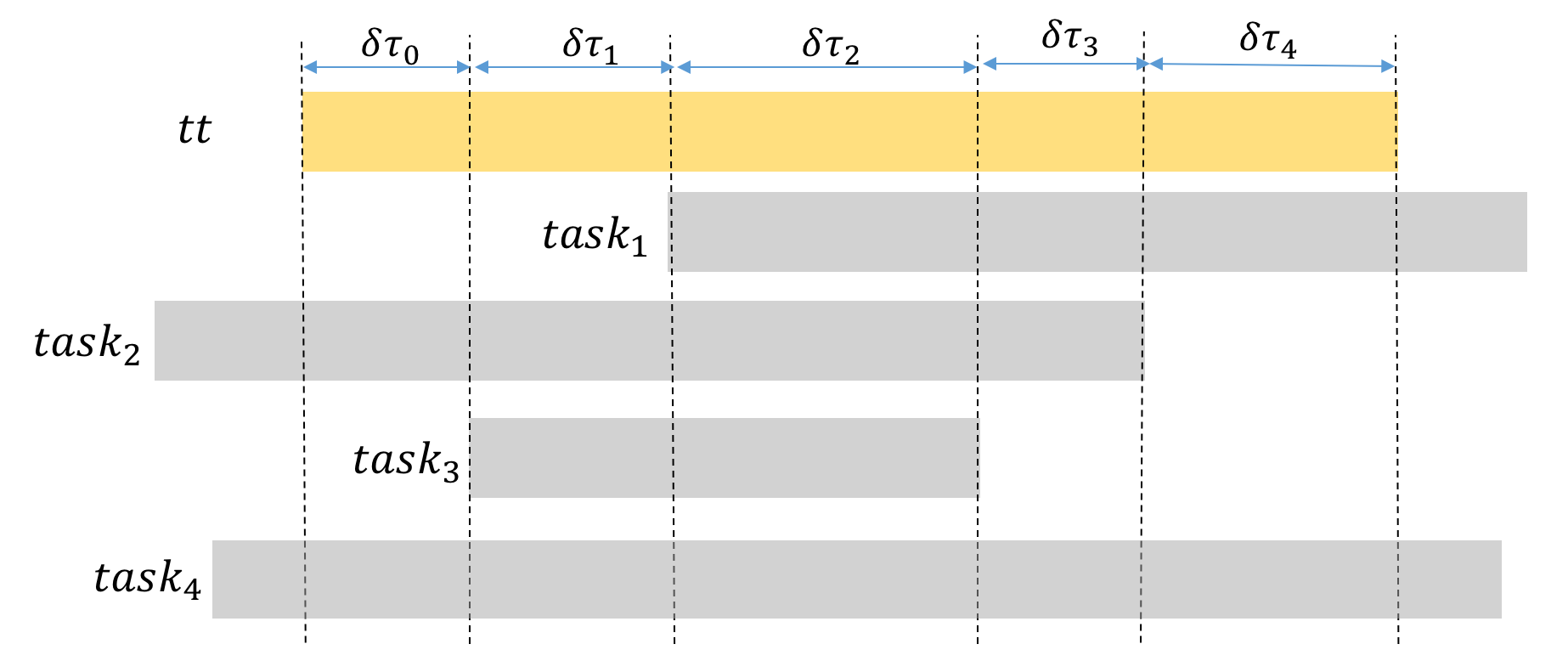}
	\caption{\textbf{\small We collect metrics data at concurrent task start and end boundaries.}}
	\label{fig:piece-wise}
	\vspace{-3mm}
\end{figure}

\vspace{-2.5mm}
For each task $tt$, we added support to collect the timeseries data for its wait-time and the corresponding data processed metrics at the boundaries of task start and finish for all other tasks $st$ concurrent to $tt$. Figure~\ref{fig:piece-wise} shows the four cases of task start end boundaries concurrent to target task $tt$. That is, instead of collecting the timeseries data at regular $\delta \tau$ intervals as shown in Figure~\ref{fig:be_theory}, we capture it only for concurrency related events that can have an impact on a task's \RATP\ values. Note that with this approach, the extent of each $\delta \tau$ depends on the frequency of arrival and exit of concurrent tasks, thus enabling us to capture the effects of concurrency on a task's execution more accurately. \par

However, if the arrival rate of concurrent tasks is low, this can affect the distributions of our metric values. To address this, we also record the metrics at heartbeat intervals in addition to above task entry and exit boundaries. On the other hand, for workloads consisting of tasks with sub-second latency, our approach gives more fine-grained windows for analysis. Section~\ref{sec:experiments} compares the impact of both the approaches on the quality of our explanations. \par

\vspace{-5mm}
\section{iQC-Graph}
\label{sec:framework}
\vspace{-2.5mm}
In the previous section, we discussed our methodology to assign blame towards a concurrent task $st$ for causing contention to a target task $tt$. As seen previously, a query typically consists of a dataflow of stages where each stage is composed of multiple parallel tasks. In order to consolidate blame caused or faced by a query, we construct \protoG, a multi-layered directed acyclic graph. \cut{. Section~\ref{subsec:levels} discusses the levels and construction of \protoG, and in Section~\ref{subsec:vc_def} we describe how we assign node weights for all vertives in the graph. We then present our blame analysis API in Section~\ref{sec:solution} that helps us meet Challenge~\ref{sub:challenge5} from Section~\ref{sec:challenges}. }

\vspace{-4mm}
\subsection{Levels in \protoG}
\label{subsec:levels}
\vspace{-2.5mm}
We now describe the levels in \protoG\ (the full pseudo-code for constructing it can be found in \cite{Proto} due to space constraints). Level 0 and Level 1 contain the target queries and \ts s respectively -- these are the queries or stages that the \admin\ wants to analyze.  On the other end of \protoG,  Level 6 represents all source queries, while Level 5 contains all \sos s, which are the queries and stages concurrently running with the target queries. Note that it is important to separate the queries and their stages in two different levels at both ends of \protoG - this is to diagnose impacts from and to multiple stages of a query. \cut{if multiple stages of a source query $Q_s$ run in parallel with multiple stages of a target query $Q_t$, $Q_s$ may have a higher impact on the performance of $Q_t$ compared to other source queries.} \par

The middle three levels -- Levels 2, 3 and 4 (described shortly) -- keep track of causes of contentions of different forms and granularity that enable us to connect these two ends of \protoG\ with appropriate attribution of responsibility to all intermediate nodes and edges. Levels 2, 3, and 4 answer the impact questions described above respectively.  \par

\cut{
\begin{table*}[ht]
	\small
	\centering
	\caption{Levels in \protoG: Levels $\ell_{2}$,$\ell_{3}$,$\ell_{4}$ represent the explanation types supported in \proto}
	\label{table:levels_in_protoGraph}
	\begin{tabular}{|p{0.6cm}||p{1.5cm}|p{3cm}|p{4.5cm}|p{6.2cm}|} \hline
		Level&Vertex Type&Description&Vertex Contribution (VC)&Comments\\ \hline \hline
		$\ell_{0}$& Target Query & Set of queries to be analyzed& 	
		\begin{equation}\label{eq:leve_0_vc}
		\VC_{u}^{\ell_{0}} = 1
		\end{equation} & - \\ \hline
		$\ell_{1}$& Target Stage & For each query at $\ell_{0}$, stages to be analyzed & 
		\begin{equation}\label{eq:leve_1_vc}
		\VC_{u}^{\ell_{1}} = WD_{ts}
		\end{equation} & cumulative CPU time (\ie, the work done $WD$) by target stage $ts$   \\ \hline
		$\ell_{2}$& Immediate Explanation & For each stage at $\ell_{1}$, its \RATP\ for each resource & 
		\begin{equation}\label{eq:leve_2_vc_tab}
		\VC_{u}^{\ell_{2}} = \sum_{tt} \int_{\tau_{tt}^{start}}^{\tau_{tt}^{end}} \RATP_{tt}^{\tau,r,h} d\tau
		\end{equation} & 
		{\small where $\tau_{tt}^{end}, \tau_{tt}^{start}$ denote the start and the end time of a task $tt$ of a \ts. $\RATP_{tt}^{\tau, r, h}$ is computed as per Equation~(\ref{eq:ratp}) given in Section~\ref{sec:blame}. } \\ \hline
		$\ell_{3}$& Deep Explanation & For each \IE\ at $\ell_{2}$, cumulative \RATP\ for tasks executed on each host & 
		\begin{equation}\label{eq:level_3_vc_tab}
		\VC_{u}^{\ell_{3}} = \sum_{tt \in h-tasks }   \int_{\tau_{tt}^{start}}^{\tau_{tt}^{end}} \RATP_{tt}^{\tau, r, h} d\tau
		\end{equation} & 
		{\small
			where, $\tau_{tt}^{end}, \tau_{tt}^{start}$ denote the start and the end time of a task $tt$ of a \ts, node $u$ corresponds to resource $r$ and host $h$, and $h-tasks$ denotes the list of tasks in this stage executing on host $h$. 	
		} \\ \hline
		$\ell_{4}$& Blame Explanation & For each \DE\ at $\ell_{3}$, blame towards each stage concurrent on the same host & 
		\begin{equation}\label{eq:level_4_vc_tab}
		\VC_{u}^{\ell_{4}} = \sum_{tt \in h-tasks } \beta_{st \rightarrow tt}^{r, h}
		\end{equation}
		&
		{\small
			where $h-tasks$ is the list of tasks executing on host $h$, and node $u$ corresponds to a source task $st$, resource $r$, and host $h$. The value of $\beta_{st \rightarrow tt}^{r, h}$ is computed as described earlier in Equation~(\ref{eq:blame_attr_uniform}) given in Section~\ref{sec:blame}.  } \\ \hline
		$\ell_{5}$& Source Stage & The {\em source stage} running concurrently with {\em target stage} & 
		\begin{equation}\label{eq:leve_5_vc}
		\VC_{u}^{\ell_{5}} = 1
		\end{equation} 
		& - \\ \hline
		$\ell_{6}$& Source Query & The corresponding {\em source query} & 
		\begin{equation}\label{eq:leve_6_vc}
		\VC_{u}^{\ell_{6}} = 1
		\end{equation} & -  \\ \hline
	\end{tabular}
\end{table*}
}
\vspace{-3mm}
\subsection{Vertex Contributions}
\label{subsec:vc_def}
For each node $u$ in the graph, we assign weights, called \textbf{V}ertex \textbf{C}ontributions denoted by $\VC_u$, that are used later for analyzing impact and generating explanations. Specifically, $\VC_u$ measures the standalone impact of $u$ toward the contention faced by any \tq\ vertex in Level-0. The \VC\ values of different nodes are carefully computed at different levels by taking into account the semantics of respective nodes. For Level 0 \tq, Level 5 \sos, and Level 6 \sq\ vertices, we currently set $\VC_u = 1$ for all nodes $u$ in each level. Our \admin\ interface allows to configure these based on the requirements of the workload, \eg\ assign higher weights to queries from certain users, or SLA-driven queries. For the Level 1 \ts\ vertices, we set this to the cumulative CPU time (\ie, the work done) by stage $s$ in order to assign higher impacts through stages that get more work done. We now describe Levels 2-4, and show how we compute these values for them.\par 

\vspace{2mm}
\noindent
\textbf{Level 2 - Track Resource Level Contentions:}
For each \ts\ in Level 1, we add five nodes in Level 2 corresponding to each of the resources, namely scheduling slots, CPU, network, IO and memory. From Equation~\ref{eq:ratp}, the $\RATP$ of a task $tt$ for resource $r$ during its execution from time $t_{start}$ to $t_{end}$ (integrate over all $\delta \tau$ intervals) on host $h$ (superscripts $r,h$ ommitted) is:
\vspace{-4.5mm}
	
\begin{equation}
\vspace{-2mm}
\RATP_{tt} = \int_{\tau_{tt}^{start}}^{\tau_{tt}^{end}} \RATP_{tt}^{\delta \tau} d\delta \tau
\end{equation}
	
For each node $u$ in Level-2, we assign its $VC_{u}^{\ell_{2}}$ to be the cumulative $\RATP$ for all tasks of this \ts\ vertex in Level-1 as: 
\vspace{-4.5mm}

\begin{equation}\label{eq:leve_2_vc}
\VC_{u}^{\ell_{2}} = \sum_{tt} \int_{\tau_{tt}^{start}}^{\tau_{tt}^{end}} \RATP_{tt}^{\tau,r,h} d\delta \tau
\end{equation} 

\begin{sloppypar} 
\noindent
\textbf{Level 3 - Track Host Level Contentions:}
Level-3 unfolds the components of wait time distribution of a \ts\ further to keep track of the $\RATP\textit{-blocked}$ values of each stage for a specific resource on each host of its execution. First, we find all the hosts that were used to execute the tasks of a particular \ts. Then, \cut{for each node in Level 2 corresponding to this \ts, we add multiple nodes in Level 3 for these hosts. Thus, }for everynode in Level 2 corresponding to resource $r$, we add $P_r \times H$ new nodes in Level 3, where $P_r$ is the number of different requests that can lead to a wait time for resource $r$, and $H$ is the number of hosts involved in the execution of this \ts. For instance, the IO bottleneck of a \ts\ can be explained by the distribution of time spent waiting for ${\scriptsize \tt IO\_READ}$ and ${\scriptsize \tt IO\_WRITE}$. Therefore, for $r$ = IO, $P_r = 2$, and for each host we add two nodes for { \tt IO\_BYTES\_READ\_RATP} and { \tt IO\_BYTES\_WRITE\_RATP} in Level 3. The Level-3 nodes for other resources are generated as discussed in Section~\ref{subsec:resources}.
 \end{sloppypar}

The value of $\VC_{u}^{\ell_{3}}$ is set similar to Equation~\ref{eq:leve_2_vc} except that the summation is done over all tasks of a \ts\ (denoted as $h-tasks$) executing only on host $h$ in question. 
\vspace{-4.5mm}

\begin{equation}\label{eq:level_3_vc}
\VC_{u}^{\ell_{3}} = \sum_{tt \in h-tasks }   \int_{\tau_{tt}^{start}}^{\tau_{tt}^{end}} \RATP_{tt}^{\tau, r, h} d\delta \tau
\end{equation} 


\vspace{2mm}
\noindent
\textbf{Level 4 - Linking Cause to Effect:}
For each node $v$ in Level 3 corresponding to a \ts, host $h$, and type of resource request $r$, if the tasks of this \ts\ were executing concurrently with tasks belonging to $P$ distinct source stages on host $h$, we add $P$ nodes $u$ in Level 4 and connect them to $v$. The $\VC_{u}^{\ell_{4}}$ of each node is then computed from the blame attribution of all target stages in Level 3 that it can potentially affect.
\vspace{-4.5mm}

\begin{equation}\label{eq:level_4_vc}
\VC_{u}^{\ell_{4}} = \sum_{tt \in h-tasks } \beta\textit{-blocked }_{st \rightarrow tt}^{r, h}
\end{equation}

\cut{We present the details of the graph construction algorithm, scalability overhead, and summarize all levels and their \VC\ computations in our technical report \cite{Proto} for a quick reference. \par}

\vspace{-1mm}
\subsection{iQC-Graph by example}
\label{subsec:iqc_by_example}
\vspace{-2.5mm}
	In example ~\ref{eg:query_dag}, suppose the user selects $Q_0$ as the target query, and wants to analyze the contention of the stages on the critical path. First, we add a node for $Q_{0}$ in Level 0, and nodes for $s_{1}, s_{3}, s_{4}, s_{5}$ in Level 1. Then in Level 2, 
	for each of these stages, the \admin\ can see five nodes corresponding to different resources. Although both $s_{1}$ and $s_{5}$ faced high contentions, using \protoG\ the \admin\ can understand questions such as whether the network contention faced by stage $s_{5}$ was higher than the IO contention faced by stage $s_{1}$, and so on. \par
	
\begin{figure}[h]
	\vspace{-3.5mm}
	\centering
	\includegraphics[height=1.8in,keepaspectratio]{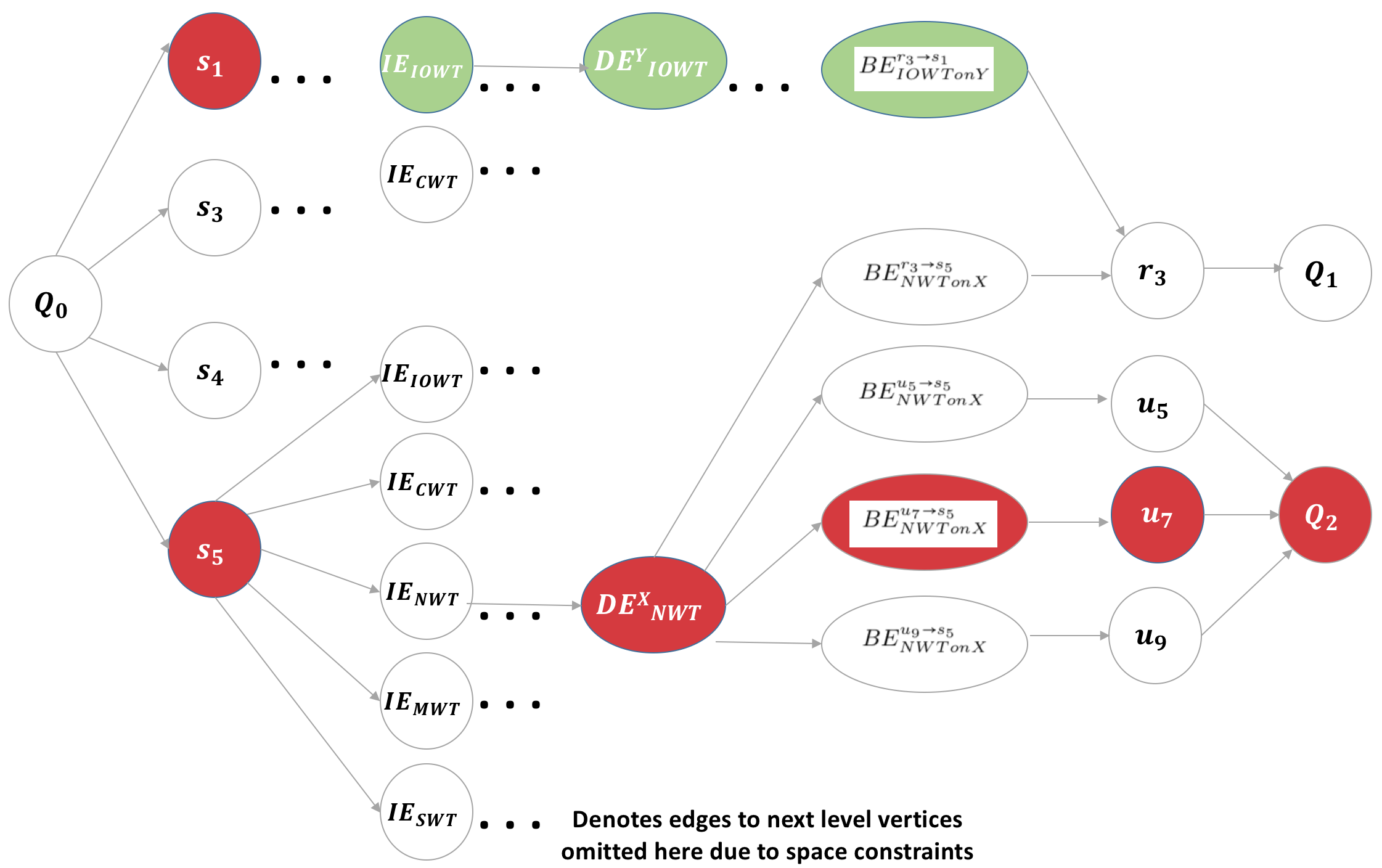}
	\caption{\textbf{\small \protoG\ illustrating the details in Example~\ref{subsec:iqc_by_example}}}
	\label{fig:graph_example}
	\vspace{-5mm}
\end{figure}

 \begin{sloppypar}	
	Suppose only the \textit{trailing} tasks of stage $s_{1}$ executing on host $Y$ faced IO contention due to data skew. Using \protoG , a deep explanation tells the user that { \tt IO\_BYTES\_READ\_\RATP} on host $Y$ for these tasks of $s_{1}$ was much less compared to the average \RATP\ for tasks of stage $s_{1}$ executing on other hosts. This insight tells the user that the slowdown on host $Y$ for tasks of $s_{1}$ was not an issue due to IO contention. If the user lists the top-$k$ nodes at Level-3 using our blame analysis API, she can see that the network \RATP\ for tasks of stage $s_{5}$ on host $X$ was the highest, and can further explore the Level-4 nodes to find the source of this disproportionality.  \par
 \end{sloppypar}
		
	Since stage $r_{3}$ of \sq\ $Q_{1}$, and stages $u_5, u_7, u_9$ of another source query $Q_2$ were executing concurrently on host $X$ with stage $s_{5}$ of $Q_{0}$, the user lists the top-$k$ Level-4 vertices responsible for this network contention for $s_{5}$. The blame analysis API outputs stage $u_7$ of query $Q_2$ as the top source of contention. Figure~\ref{fig:graph_example} shows some relevant vertices for this example to help illustrate the levels in \protoG. 
	
\vspace{1mm}
\noindent
\textbf{Discussion:} 
\cut{Once we capture the individual task-level values for the wait time distributions, we compute the cumulative time a stage has spent blocked for resource $r$, \ie\ $ \tau_{s} = \sum_{t \in s}^{ } \tau_{t} $, where $\tau_{t}$ is the wait-time of task $t$ for resource $r$. For example, \NWT\ of a stage refers to the cumulative time spent blocked on network by all tasks in that stage. } Note that we consider the cumulative \RATP\ values for computing stage-level values (in Equation~\ref{eq:leve_2_vc} and Equation~\ref{eq:level_3_vc}). The other alternative approaches include considering \textit{max} or \textit{average} values for the tasks in a stage; however, \textit{sum} captures the \textbf{total cluster time} (similar to the notion of \emph{Database Time} in \cite{OracleADDM}) spent on waiting for resources per unit data by each stage and, thus, enables us to analyze the overall slowdown of a stage in the cluster. \cut{The problem with using \textit{max} for our requirement is that if the tasks of a single stage of a \sq\   }

\cut{This enables us to compare disproportionalities in the wait-times of two stages (or queries) irrespective of their degree of parallelism in the cluster.}

\vspace{-4mm}
\section{Explanations}
\label{sec:solution}
In this section, we describe how the choice of a graph-based model enables us to consolidate the contention and blame values for generating  explanations for a query's peformance. 

\vspace{-3mm}
\subsection{Generating Explanations}\label{sec:proto_metric}
\label{sub:metrics_for_eval}
The Explanations Generator module in \proto\ uses the \VC\ values to update two metrics, namely Impact Factor (\IF) and Degree of Responsibility (\DOR). We set the \IF\ as edge weights and \DOR\ as the node properties. \par

\begin{figure}[h]
	\vspace{-3mm}
	\centering
	\includegraphics[scale=0.22,keepaspectratio]{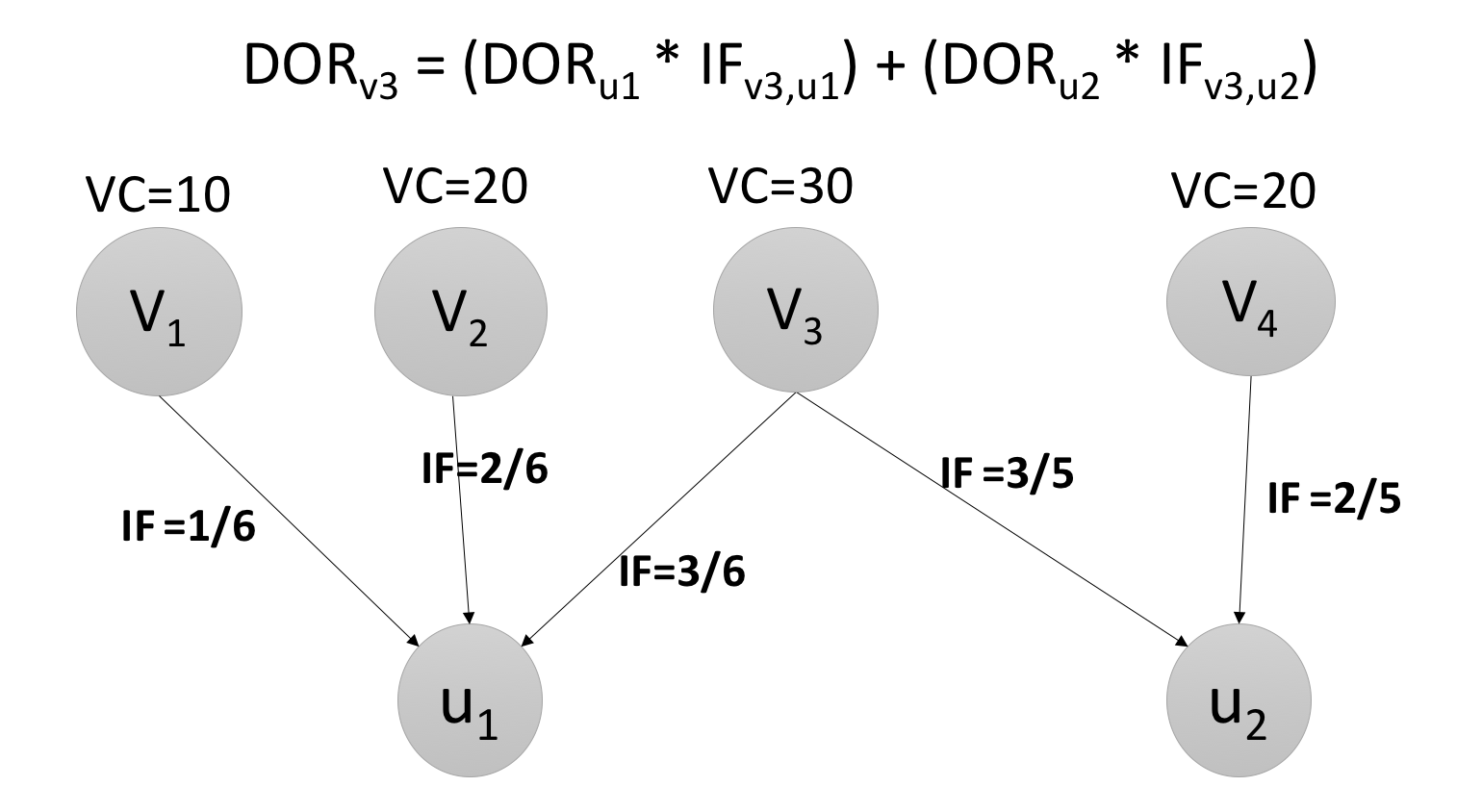}
	\caption{{\bf \small Example computation of \DOR\ from \IF\ and \VC\ values.}}
	\label{fig:weights}
	\vspace{-2mm}
\end{figure}

\vspace{2mm}
\noindent
\textbf{Impact Factor (\IF):}
Once the Vertex Contributions $\VC_u$ of every node $u$ in \protoG\ is computed to estimate the standalone impact of $u$ on a \ts, we compute the Impact Factor $\IF_{uv}$ on the edges $(u, v)$. This enables us to distribute the overall impact received by each child node $v$ among its parent nodes $u$-s. For instance, $\IF_{uv}$ from a \DE\ node $u$ to an \IE\ node $v$ gives what fraction of total impact on an \IE\ node can be attributed to each of its parent \DE\ nodes. Figure~\ref{fig:weights} shows an example of the impact received by node $u_{1}$ from nodes $v_{1}, v_{2}, v_{3}$.

The \IF\ values of the edge weights are used to generate explanations as follows:
\begin{sloppypar}
		An {\em \bf explanation} in \proto\ takes the form: \\
		Expl($~ \targetquery, ~ \targetstage, ~ \resource, ~ \subresource,  ~ \host, ~ \sourcestage, ~ \sourcequery, ~ \pathimpact$)\\
		where ,\\ 
		\targetquery\ denotes the target stage being explained by $\expl$;
		\targetstage\ denotes the target stage being explained by $\expl$;
		\resource\ $\in {CPU, Memory, IO, Network, Scheduling}$;
		\subresource\ is the type of resource impacted (see Section~\ref{subsec:resources});
		\host\ is the host of impact; 
		\sourcestage\ is the stage of the concurrent query that has caused impact;
		\sourcequery\ is the impacting source query;
		$\pathimpact$ is the cumulative weight of the path originating from \sourcequery\ and ending at \targetquery.
\end{sloppypar}

\vspace{-3mm}
\subsection{Contention Analysis using iQC-Graph}
\vspace{-1.5mm}
A top-$k$ analysis lets user specify the number of top explanations she is interested in. Additionally, \proto\ consolidates the explanations further to 
aggregate responsibility of each entity (queries, stages, resources and hosts) towards the contention faced by every \tq. 

\vspace{2mm}
\noindent
\textbf{Degree of Responsibility (\DOR):}
Finally we compute the Degree of Responsibility $\DOR_u$ of each node $u$, which stores the \emph{cumulative impact} of $u$ on a \tq\ $t$. The value of $\DOR_u$ is computed as the \emph{sum of the weights of all paths} from any node $u$ to the \tq\ node $t$, where the weight of a path is the product of all $\IF_{vw}$ values of all the edges $(v, w)$ on this path. If we choose more than one query at Level 0 for analysis, a mapping of the values of \DOR\ toward each query is stored on nodes at Level 5 and 6. The computation of \DOR\ of node $v_{3}$ is illustrated in Figure ~\ref{fig:weights}. It can be noted that since our framework incorporates impacts originating from the synthetic `Unknown' source, it enables an \admin\ to rule out slowdown due to concurrency issues if the impact through this nodes is high. \cut{We give the details of its computation, the algorithm, and runtime analysis in Appendix~\ref{app:algos_dor}.} Using \DOR, users can perform the following use-cases that enable deep exploration of contentions in a shared cluster:

\begin{itemize}[leftmargin=*]
	\itemsep0em
	\item \textbf{Finding Top-K Contentions for Target Queries:} For any query in the workload, users can find answers for: (a) What is the impact through each resource for its slowdown? (b) What is the impact through each host for a specific resource? and, (c) What is the impact through each \sq\ or \sos\ on a specific host and resource combination?
	\item \textbf{Detecting Aggressive Source Queries:} \proto\ allows the \admin\ to do a top-down analysis on a \sq\ or \sos\ to explore how it has caused high impact to all concurrent queries. 
	To detect such aggressive queries, we 
	find the (top-$k$) Level 6 nodes having the highest value of total \DOR\ toward all affected queries 
	(recall that for each source query, we keep track of the \DOR\ value toward each target query). 
	\item \textbf{Identifying Slow Nodes and Hot Resources:} Performing a top-$k$ analysis on levels 2 (resources) and 3 (hosts) will yield the hot resource and its corresponding slow node with respect to a particular \tq. In order to get the overall impact of each resource or each host on all target queries, \proto\ provides an API to (i) \emph{detect slow nodes}, \ie, group all nodes in Level 3 by hosts, and then output the total outgoing impact (sum of all \IF\ values) per host, and (ii) \emph{detect hot resources}, \ie,  - output the total outgoing impact per wait-time component nodes in Level 2.
\end{itemize}

\cut{
\subsection{Rules for Cluster Scheduler}
\label{subsec:rules}
Our ongoing research adds an \textit{actuator engine} to \proto\ where a set of rules is recommended using the blame analysis API on \protoG\ to be applied by a cluster scheduler in an \textit{online} execution of workloads. Here, we present an example of two types of rules that can be used by a cluster scheduler for recurring executions of workloads: (a) alternate query placement and (b) query priority readjustment. \cut{Our rule generator uses the blame analysis API described in previous section to output Json rules. To test their efficacy, we extended the Spark scheduler to parse and apply a selected rule from this file.} \par

\cut{
\begin{sloppypar}
\begin{definition} \label{def:rule}
	A {\em \bf rule} $\prule$ in \proto\ is a 6-tuple: \\
	$\prule = \langle \id,$ $\pruletype,$ $\pquery,$ $\pquerytype,$ $\pvalue,$ $\porder \rangle$, 
	where  
	\id\ is the unique identifier of $\prule$,
	\pruletype\ is one of the types discussed below,
	\pquery\ is the query for which the rule is applicable,
	\pquerytype\ denotes whether it is a source or target query,
	\pvalue\ is used by the scheduler to quantify the rule based on the \pruletype,
	\porder\ is the order in which to apply rules of the same \pruletype.
\end{definition}
\end{sloppypar}
}
	
	\textbf{(1) Alternate Query Placement ($\prule -AGG$):~} In this type of rules, we output the top-$k$ aggressive queries identified by \proto\ using the approach presented in Section~\ref{subsec:aggressive}. The \pquery\ is the \id\ for this query, and \pquerytype\ is set as `Source'. Finally, the \pvalue\ is set to be the maximum share this \pquery\ was entitled to in a \FAIR\ allocation policy, \ie\ $\frac{1}{n}$ where $n$ is the number of maximum query concurrency in the workload. Our extension to Spark scheduler takes the rule \id\ as input and if it matches the type $\prule -AGG$, it places the corresponding query in a dedicated pool with its share set to \pvalue. 
	\par
	\textbf{(2) Query Priority Readjustment ($\prule -DYNP$):~} The second type of rules generates two types of priority rectification suggestions: (i) for the top-$k$ affected target queries that suffered the highest impact, we generate $k$ rules for each \pquery. The \pvalue\ is then set to $\pvalue = P_{orig} + i$ where $P_{orig}$ is the priority of the corresponding \pquery\ in the previous workload, and $i$ is the rank of the queries output in the top-$k$ analysis. Owing to a priority higher than its previous execution, each of the query's tasks get more opportunity to be launched. Similarly, (ii) for top-$k$ impacting source queries, the rule outputs new priorities that are lower than the priority each query had in its previous execution.  

Currently, \proto\ applies the rules selected by the user. \cut{Moreover, we currently do not support any ranking or ordering of the rules apart from simply assigning the rank output by \proto 's top-$k$ analysis API\footnote{Research to infer a ranking for rules is an on-going effort.} since the focus of this work is in providing a framework for blame attribution and demonstrating the potential of our model for generating prescriptions as well. }
We show through our experiments in Section~\ref{sec:experiments} how application of these rules creates interventions in the original schedule that benefit the affected queries. 
}

\vspace{-8mm}
\section{Evaluation}
\label{sec:experiments}
\vspace{-3mm}
Our experiments were conducted on Apache Spark 2.2 \cite{Spark} deployed on a 20-node local cluster (master and 19 slaves). Spark was setup to run using the standalone scheduler in \FAIR\ scheduling mode \cite{DelayScheduling} with default configurations. Each machine in the cluster has 8 cores, 16GB RAM, and 1 TB storage. A 300 GB TPC-DS \cite{TPCDS} dataset was stored in HDFS and accessed through Apache Hive in Parquet \cite{Parquet} format. The SQLs for the TPC-DS queries were taken from the implementations in \cite{sparksqlperf} without any modifications. \par

\textbf{Workload:} Our analysis uses a TPCDS benchmark workload that models multiple users running a mix of data analytical queries in parallel. We have 6 users (or tenants) submitting their queries to their dedicated queue. Each user runs queries sequentially from a series of randomly chosen 15 TPCDS queries. When submitting to a queue, the selection of queries is randomized for each user. The query schedules were serialized and re-used for each of our experiments to enable us to compare and contrast results across executions. During the experiments, our cluster slots were 97\% utilized and Ganglia showed around 36\% average CPU utilization.  \par

We now describe how \proto\ enables deeper diagnosis of contentions compared to (a) \textbf{Blocked-Time Analysis (BTA)}: we use the values of blocked times for IO and Network~\cite{BlockedTime} and aggregate them at stage and query levels, (b) \textbf{Naive-Overlap}: based on overlap times of query runtimes (a technique popularly used by support staff trying to resolve \textit{who is affecting my query} tickets), and (c) \textbf{Deep-Overlap}: we compute the cumulative overlap time between all tasks of a pair of concurrent queries.

\begin{figure}[h]
	\vspace{-2.5mm}
	\centering
	\begin{subfigure}{.26\textwidth}
		\includegraphics[width=1.7in,keepaspectratio]{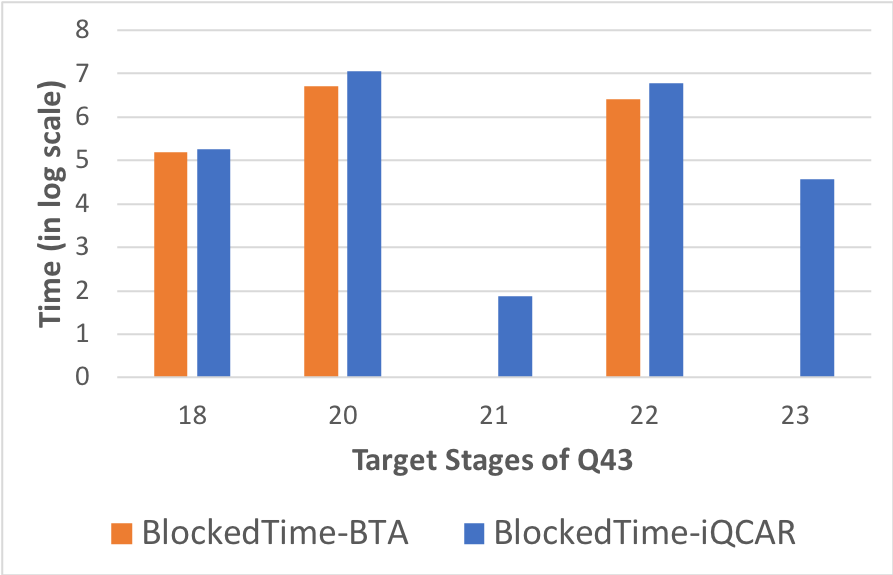}
		\caption{}
		\label{fig:blockedTimeCompare}
	\end{subfigure}%
	\begin{subfigure}{.26\textwidth}
		\includegraphics[width=1.7in,keepaspectratio]{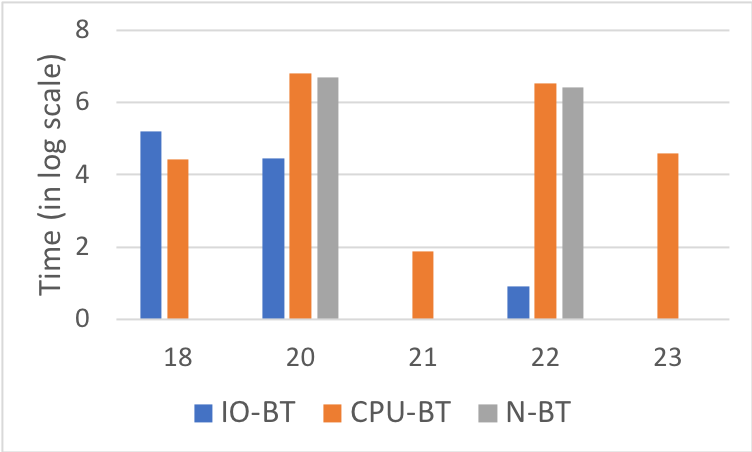}
		\caption{}
		\label{fig:bt_compare}
	\end{subfigure}%
	\caption{\textbf{\small  (a) Resource blocked times from \proto\ and \textbf{BTA} approach; (b) Blocked times for IO, CPU and Network.}}
	\label{fig:bta}
	\vspace{-3.5mm}
\end{figure}

\vspace{-3.5mm}
\subsection{Comparison with Baseline}
\label{subsec:baseline}
\begin{figure}[h]
	\vspace{-2.5mm}
	\centering
	\begin{subfigure}{.25\textwidth}
		\includegraphics[width=1.5in,keepaspectratio]{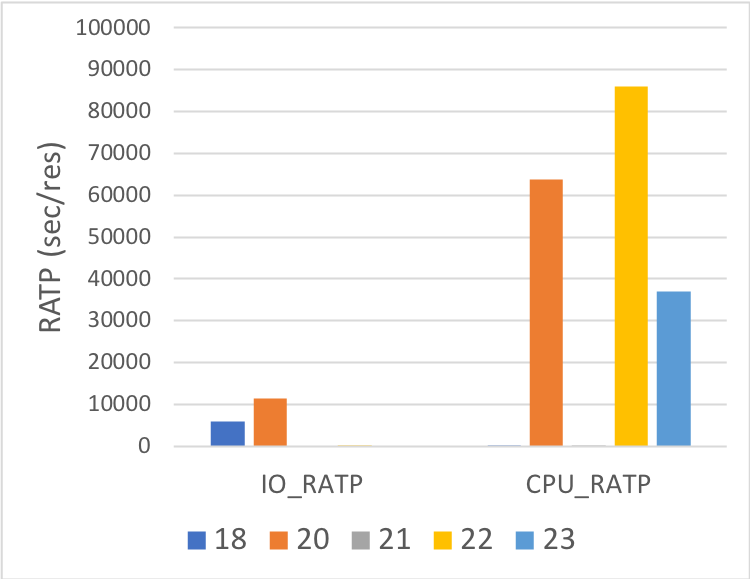}
		\caption{}
		\label{fig:ratp_compare}
	\end{subfigure}%
	\begin{subfigure}{.25\textwidth}
		\includegraphics[width=1.5in,keepaspectratio]{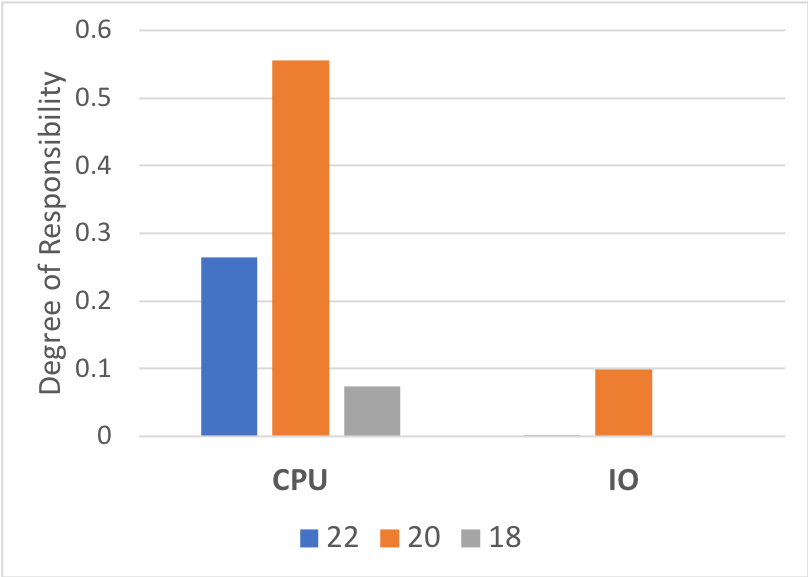}
		\caption{}
		\label{fig:dor_on_stages}
	\end{subfigure}%
	\caption{\textbf{\small (a) \RATP\ for IO and CPU for high-impacting stages, and (b) their \DOR\ in impacting $Q_{43}$. }}
	\label{fig:iqcar}
	\vspace{-3.5mm}
\end{figure}

For this experiment, we identify a \tq\ $Q_{43}$ from our workload that was 178\% slower compared to its unconstrained execution (when run in isolation). Figure~\ref{fig:blockedTimeCompare} compares the bottlenecks observed (y-axis is on log scale) with the \textit{BTA} method vs the blocked times captured by \proto\ for its impacted stages (other stages had minimal blocked times to compare). In addition to the blocked times from network and IO, \proto\ accrues additional blocked times that help to identify potential causes of slowdown for stages like 21 and 23. Figure~\ref{fig:bt_compare} compares the relative per resource blocked times. The network and CPU blocked time is significantly high for stages 20 and 22. However, when we compare the \RATP 's for these stages using \proto, we see in Figure~\ref{fig:ratp_compare} that the CPU \RATP\ was much higher for multiple stages compared to their IO \RATP, whereas the network \RATP\ was significantly low to even compare. If we analyze the impact on each of these stages generated using the explanations module, stage 20 received the maximum impact through CPU compared to others. This detailed analysis cannot be obtained from looking at \textit{BTA} alone. \par     

\begin{figure}[t]
	\centering
	\begin{subfigure}{.2\textwidth}
		\includegraphics[width=1.3in,keepaspectratio]{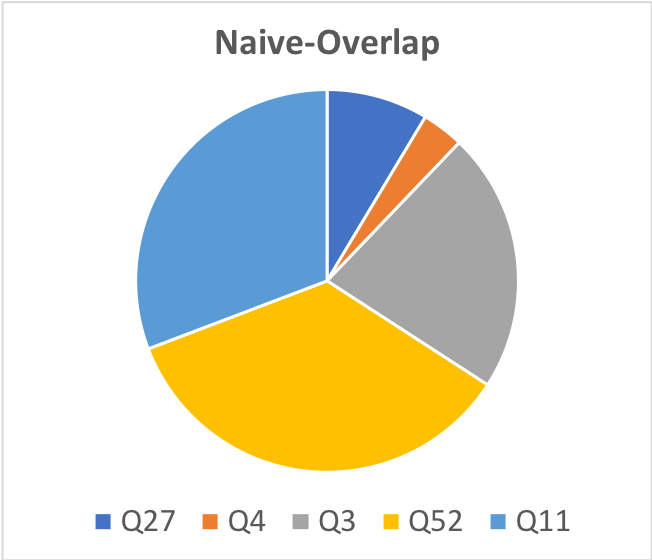}
		\caption{}
		\label{fig:baseline-naive-overlap}
	\end{subfigure}%
	\begin{subfigure}{.2\textwidth}
		\includegraphics[width=1.3in,keepaspectratio]{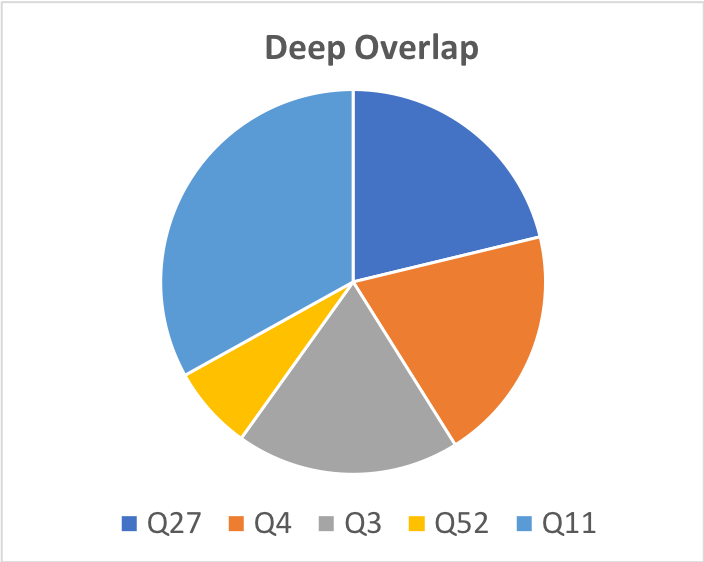}
		\caption{}
		\label{fig:baseline-deep-overlap}
	\end{subfigure}%
	\begin{subfigure}{.2\textwidth}
	\includegraphics[width=1.1in,keepaspectratio]{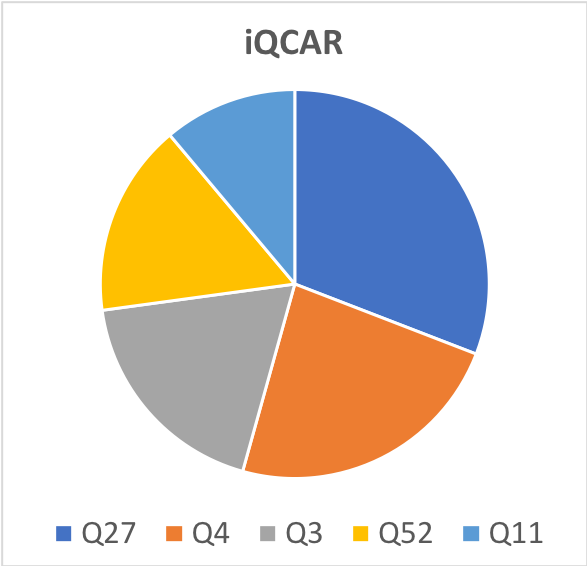}
	\caption{}
	\label{fig:baseline-iqcar}
\end{subfigure}%
	\caption{\textbf{\small Compare top-$5$ impacting queries between (a) \textit{Naive-Overlap} (b) \textit{Deep-Overlap} and (c) \proto . }}
	\label{fig:baseline}
	\vspace{-4.5mm}
\end{figure}

Next, for assessing who is causing the above slowdown to $Q_{43}$, we compare our results with \textit{Naive-Overlap} and \textit{Deep-Overlap}. For the overlap-based approaches, highest blame is attributed to query with most overlap. It can be seen from Figure~\ref{fig:baseline} that the top queries that overlap differ significantly between \textit{Naive-Overlap} and \textit{Deep-Overlap}. The queries with minimal overlap shown in \textit{Naive-Overlap} ($Q_4$  and $Q_{27}$) have relatively more tasks executing in parallel on the same host as that of target query, causing higher cumulative \textit{Deep-Overlap}. The output from \proto\ is more comparable to \textit{Deep-Overlap}, but has different contributions. Especially for $Q_{11}$, where the tasks had a high overlap with tasks of our \tq\ $Q_{43}$, the impact paths showed low path weights between these end points. A further exploration revealed that only 18\% of the execution windows (captured via the time-series metrics), showed increments in CPU and IO acquired values for $Q_{11}$ for the matching overlapping windows. $Q_{11}$ itself was blocked for these resources in 64\% of the overlapping windows. Without the impact analysis API of \proto, an \admin\ will need significant effort to unravel this. \par

\vspace{-5.5mm}
\subsection{Test Cases}	 
\vspace{-1.5mm}
The purpose of our second set of experiments is three-fold to: (i) verify that \proto\ detects symptoms of contentions induced in our workload and outputs newly identified sources, (ii) demonstrate how \proto\ achieves this for a running workload, and (iii) how we can perform a time-series analysis with \proto. Our test-cases for induced resource contention scenarios are summarized in Table~\ref{table:exp_contentions}. \proto\ detects contentions for Spark's managed memory, hence the test-case of Memory-external is out of scope. Note, we were not able to isolate contention only due to increased shuffle data as it was always resulting in increasd IO contention too owing to shuffle write. So induction of network contention in our workload (sql queries) was not possible. We now share our induction methodology and observations:

\begin{table*}[ht]
	\vspace{-2.5mm}
	\small
	\centering
	\caption{Summary of Induced Contention Scenarios}
	\label{table:exp_contentions}
	\vspace*{-3mm}
	\begin{tabular}{|p{2.8cm}||p{14cm}|} \hline
		Test-Case & Detail \\ \hline \hline
		CPU-Internal & User-7 submits a CPU-intensive custom SparkSQL query just \textit{after} $Q_{43}$ starts.  \\ \hline
		CPU-External & We inject a CPU-intensive custom query in a seperate Spark application instance.  \\ \hline
		IO-Internal & User-7 submits a IO-intensive custom SparkSQL query at a random time.  \\ \hline
		IO-External & A large 60GB file is read in 256MB block size on every host just \textit{after} $Q_{43}$ starts.  \\ \hline
		Memory-Internal & User-7 caches \textit{web\_sales} table just \textit{before} $Q_{43}$ starts.  \\ \hline
	\end{tabular}
	\vspace{-2mm}
\end{table*}

\vspace{-3.5mm}
\subsubsection{CPU-Internal} 
\vspace{-1.5mm}
For this experiment, we use the same workload as before (the schedules were serialized), and induce contention query $Q_{cpu-int}$ shown in Listing~\ref{lst:indc_cpu_int} along with our \tq\ $Q_{43}$. This query $Q_{cpu-int}$ exhibits the following characteristics: (a) low-overhead of IO read owing to a scan over two small TPCDS \textit{customer} tables each containing 5 million records stored in parquet format, (b) low network overhead since we project only two columns, (c) minimal data skew between partitions as the join is on \textit{c\_customer\_sk} column which is a unique sequential id for every row and shuffle operation used  hash-partitioning , and (d) high CPU requirements owing to the \textit{sha2} function on a long string (generated by using the repeat function on a string column). We now analyze the contentions of \tq\ $Q_{43}$ further. Note that our induction can affect $Q_{43}$ only to a certain extent owing to multiple aspects of scheduling on a cluster, for example the concurrency of tasks of $Q_{cpu-int}$ with tasks of $Q_{43}$ is dependent on the number of slots it can get under the configured scheduling policy. Despite this, \proto\ is able to capture $Q_{cpu-int}$ as one of the top sources in the concerned window as we show next. \par

\vspace{-1.5mm}
\begin{lstlisting}[caption={$Q_{cpu-int}$: \textit{CPU-Internal} Induction query}\label{lst:indc_cpu_int},language=SQL]
with temp 
(select c1.c_first_name as first_name, 
	sum(sha2(
		repeat(c1.c_last_name,
		45000),512)) as shasum 
	from customer c1,customer c2 
	where c1.c_customer_sk 
		= c2.c_customer_sk 
	group by c1.c_first_name) 
select max(shasum) 
from temp 
limit 100;
\end{lstlisting}

\begin{figure}[h]
	\vspace{-2.5mm}
	\centering
	\includegraphics[height=1.5in,keepaspectratio]{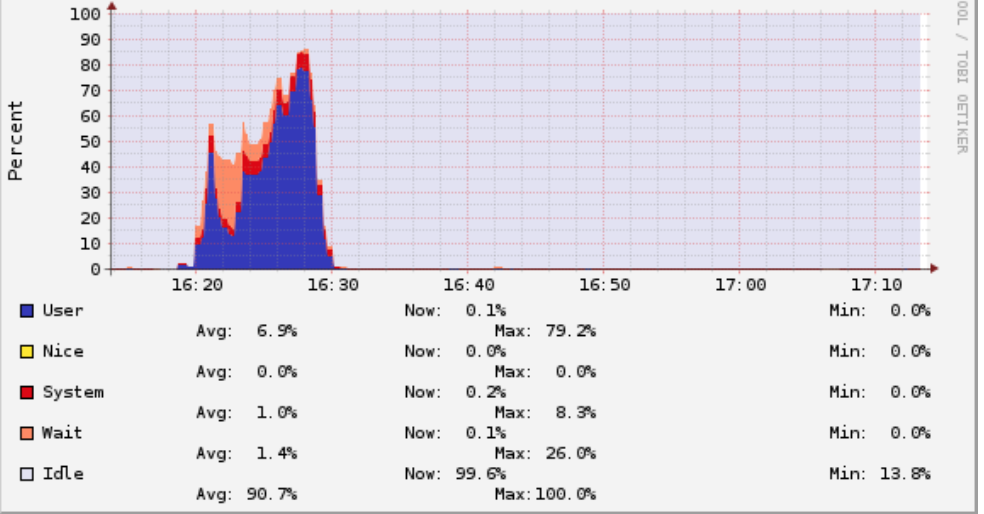}
	\caption{\textbf{\small Ganglia snapshot showing CPU Utilization when $Q_{cpu-int}$ was running. }}
	\label{fig:ganglia_indc_cpu_int}
	\vspace{-3.5mm}
\end{figure}

\noindent
\textbf{Observations:} 
Figure~\ref{fig:ganglia_indc_cpu_int} shows that the CPU utilization reaches almost 80\% during our induction, much higher than the 36\% average utilization for our workload. 

\begin{figure}[h]
	\vspace{-2.5mm}
	\centering
	\includegraphics[height=1.3in,keepaspectratio]{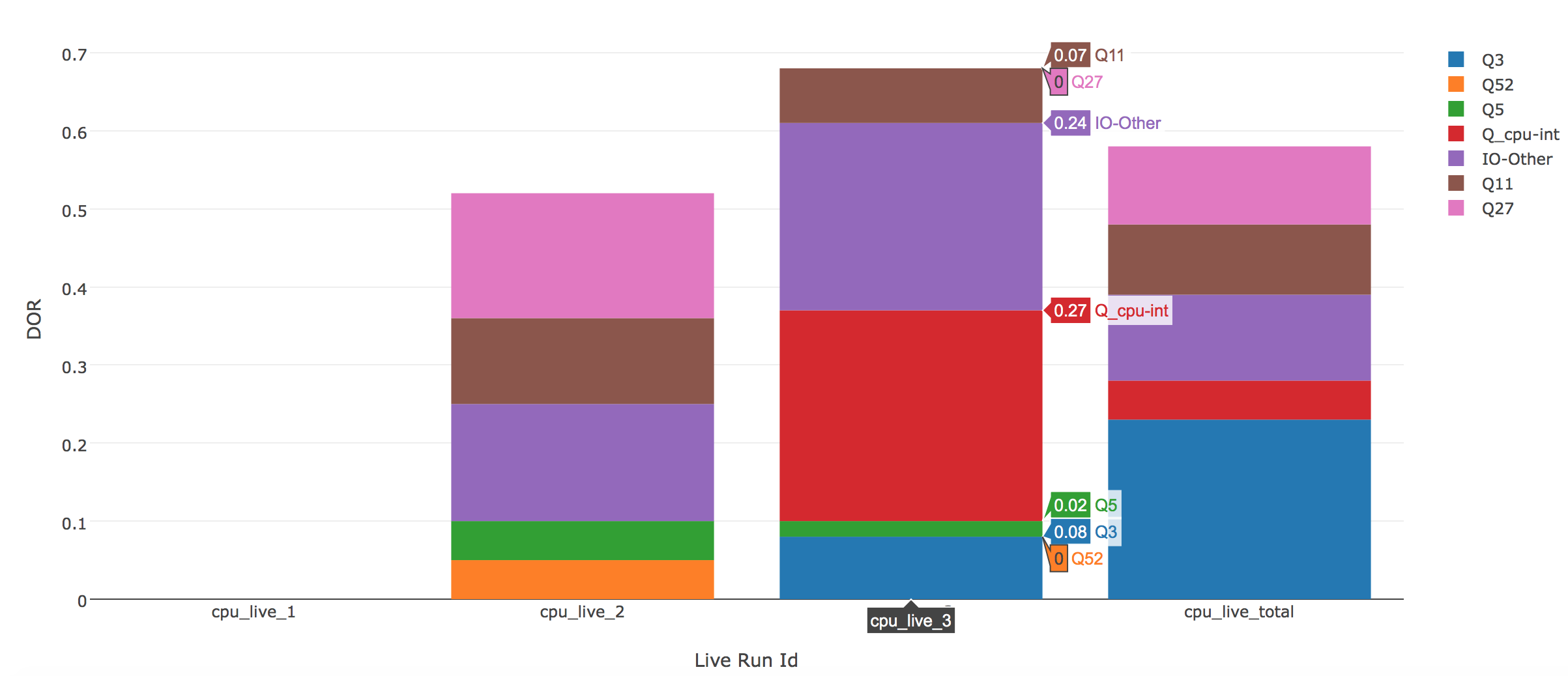}
	\caption{\textbf{\small \DOR\ of $Q_{cpu-int}$ is about 27\% towards query $Q_{43}$ during CPU induction. }}
	\label{fig:indc_cpu_dor}
	\vspace{-3.5mm}
\end{figure}

Figure~\ref{fig:indc_cpu_dor} shows the change in \DOR\ values of the source queries towards $Q_{43}$. The live analysis windows \textit{cpu\_live\_1}, \textit{cpu\_live\_2}, \textit{cpu\_live\_3} and \textit{cpu\_live\_total} show the periods in which we invoke \proto\ to analyze contentions on $Q_{43}$. The stacked bars show the relative contribution from each of the top-$5$ running sources, and their heights denote the total contribution from these five sources. Since we are analyzing contentions on only $Q_{43}$, the \textit{cpu\_live\_1} window shows no contributions as $Q_{43}$ had not started yet. $Q_{cpu-int}$ was induced at the end of window \textit{cpu\_live\_2}, hence no contribution from $Q_{cpu-int}$ in this window. However, as seen in \textit{cpu\_live\_3}, once we induce $Q_{cpu-int}$ when $Q_{43}$ starts, its contribution of 27\% is detected. 

\begin{figure}[h]
	\vspace{-2.5mm}
	\centering
	\begin{subfigure}{.25\textwidth}
		\includegraphics[height=1.3in,keepaspectratio]{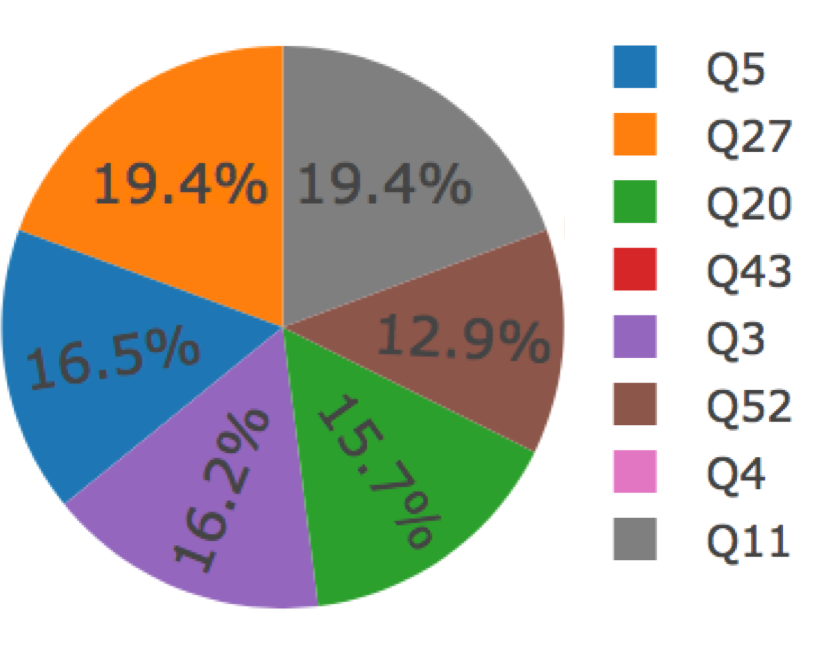}
		\caption{Naive Overlap}
		\label{fig:naive_overlap_indc_cpu}
	\end{subfigure}%
	\begin{subfigure}{.25\textwidth}
		\includegraphics[height=1.3in,keepaspectratio]{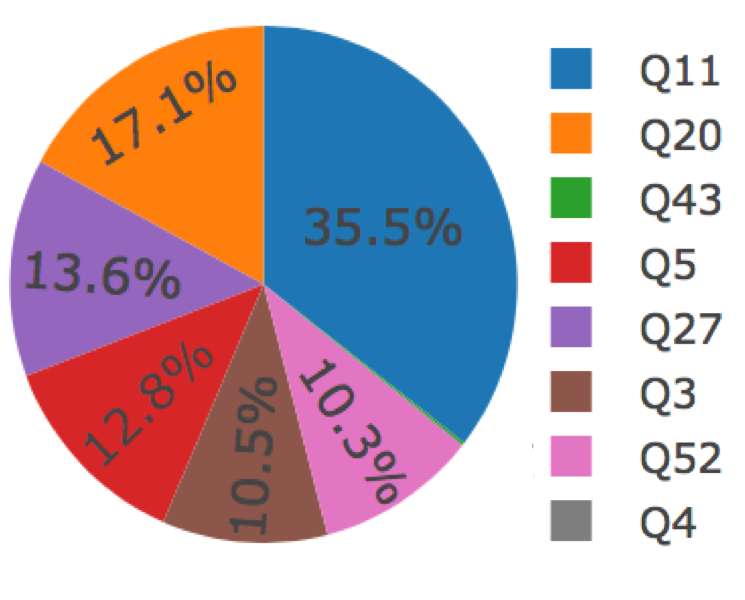}
		\caption{Deep Overlap}
		\label{fig:deep_overlap_indc_cpu}
	\end{subfigure}%
	\caption{\textbf{\small Impact from source queries. }}
	\label{fig:overlap_indc_cpu_int}
	\vspace{-3.5mm}
\end{figure}

The end-of-window analysis in \textit{cpu\_live\_total} shows the overall impact $Q_{43}$ from all sources during its execution. Although $Q_{cpu-int}$ caused high contention to $Q_{43}$ for a period, its overall impact was still limited ($0.05\%$). Details like this are missed by alternative approaches like \textit{Naive-Overlap}) and \textit{Deep-Overlap} as shown in Figure~\ref{fig:overlap_indc_cpu_int}. Moreover, an analysis with even \textit{Deep-Overlap} can mislead the user into attributing about 35\% of received impact to $Q_{11}$, whereas, the actual impact output by \proto\ shows less than 1\% overall impact from $Q_{11}$. \par

\subsubsection{IO-External} 
To induce IO outside of Spark, we read and dump a large file on every host in the cluster after the workload stabilizes (at least one query is completed for each user). Let us call this query as $IO-Other$. The timing of induction was chosen such that it overlaps with the scan stage of $Q_{43}$. We created a 60GB file on each host and used the command in~\ref{lst:indc_io_ext} to read in blocks of size 256MB using the below command:

\begin{lstlisting}[caption={$IO-Other$: \textit{IO-External} Induction query}\label{lst:indc_io_ext},language=BASH]
dd if=~/file_60GB of=/dev/null bs=256
\end{lstlisting} 

\begin{figure}[h]
	\vspace{-2.5mm}
	\centering
		\includegraphics[height=1.5in,keepaspectratio]{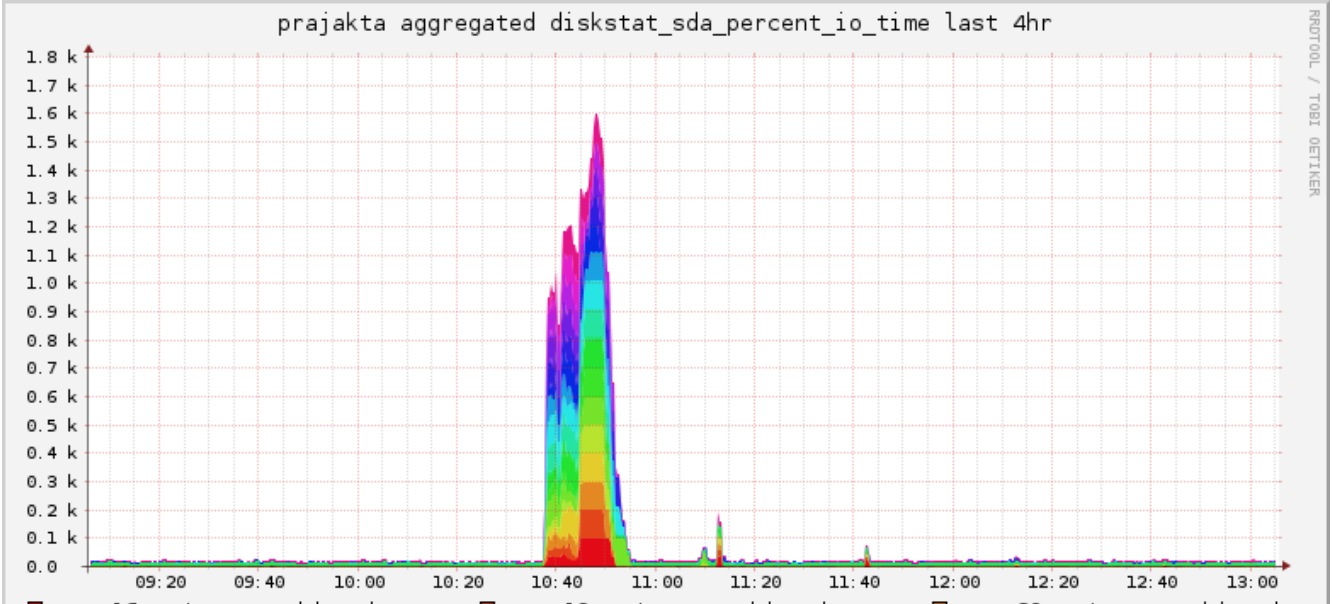}
	\caption{\textbf{\small Ganglia snapshot showing aggregate IO utilization during our external induction. }}
		\label{fig:ganglia_indc_io_ext}
	\vspace{-3.5mm}
\end{figure}

\begin{figure}[h]
	\vspace{-2.5mm}
	\centering
		\includegraphics[height=1.3in,keepaspectratio]{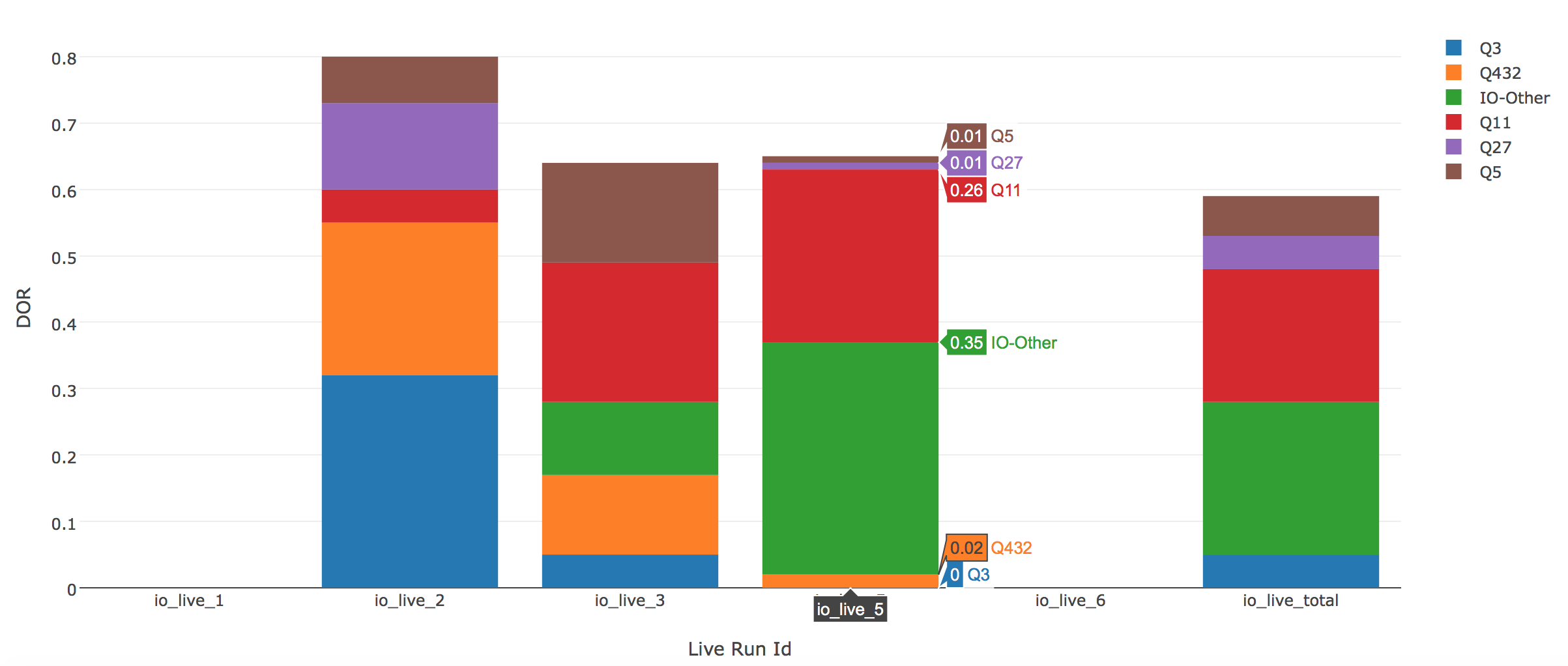}
\caption{\textbf{\small \DOR\ of source queries towards query $Q_{43}$ changes as the workload progresses. }}
		\label{fig:indc_io_ext_dor}
	\vspace{-3.5mm}
\end{figure}

\noindent
\textbf{Observations:} Figure~\ref{fig:ganglia_indc_io_ext} shows over 1600\% aggregate disk utilization for all nodes (19 slaves) in the cluster during this period of contention. We analyze impacts from sources for the following windows: (a) $Q_{43}$ had not started in \textbf{io\_live\_1}, (b) \textbf{io\_live\_2} analysis was done for \tq\ $Q_{43}$ just before we induced $IO-Other$, (c) \textbf{io\_live\_3} to \textbf{io\_live\_5} windows while $IO-Other$ is running concurrently with $Q_{43}$ (we omit showing \textbf{io\_live\_4} due to plotting space constraints), (d) $Q_{43}$ finishes execution before \textbf{io\_live\_6} ,and (e) the \textbf{io\_live\_total} window to analyze overall impact on $Q_{43}$ from the beginning of the workload till the end.  Figure~\ref{fig:indc_io_ext_dor} shows the relative contributions from each of the concurrent sources, showing that \proto\ detects the induced contention first in \textbf{io\_live\_3} (shown in green), and outputs an increasing impact during \textbf{io\_live\_4} and \textbf{io\_live\_5} when it peaks. \cut{Note how \proto\ also detects IO contention due to other external processes even before we induce our $IO-Other$ query, although the contribution was relatively much lower. }

\vspace{-2.5mm}
\subsubsection{Mem-Internal} 
Since we monitor contention only for the managed memory within Spark, our internal memory-contention test-case caches a large TPCDS table (\textit{web\_sales}) in memory just before $Q_{43}$ is submitted. Let's call this query as $Q_{mem-int}$. \par

\begin{figure}[h]
	\vspace{-2.5mm}
	\centering
	\includegraphics[height=1.3in,keepaspectratio]{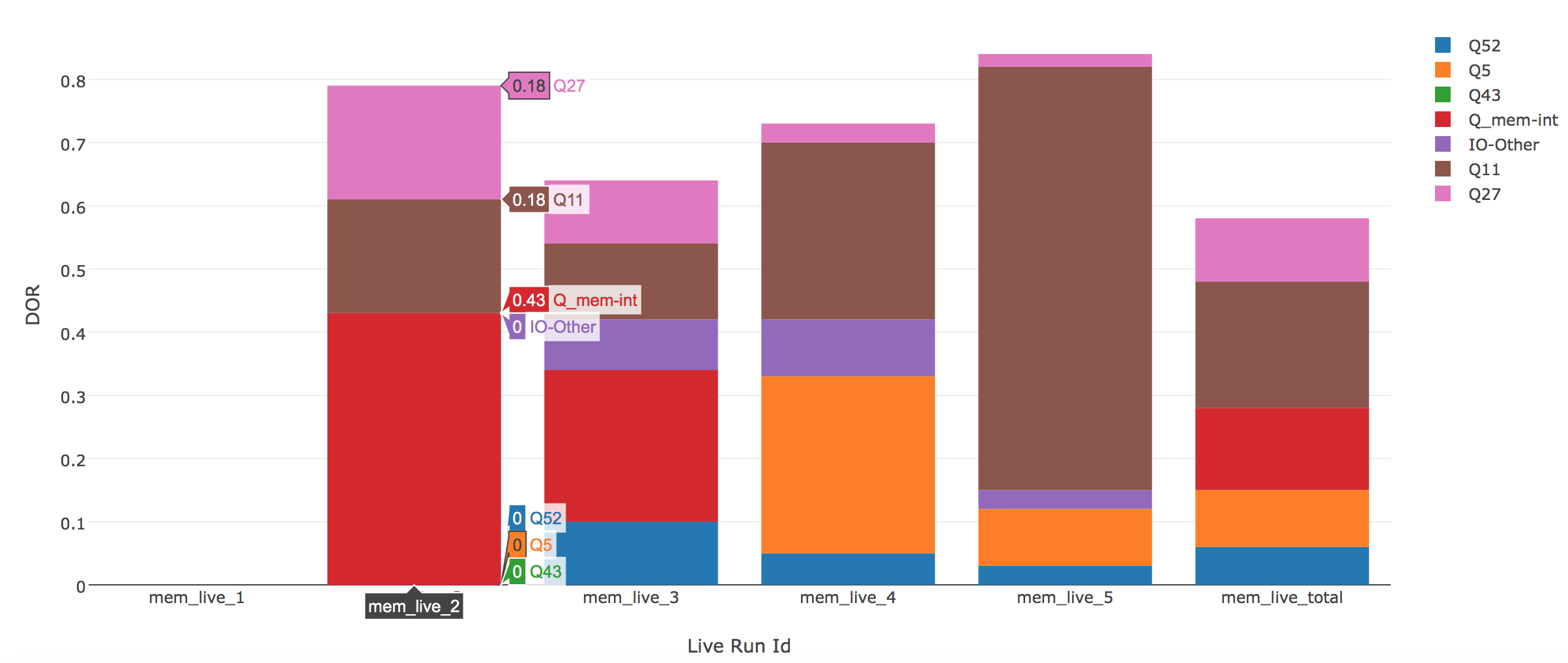}
	\caption{\textbf{\small Impact of induced internal memory contention on $Q_{43}$. }}
	\label{fig:indc_mem_dor}
	\vspace{-3.5mm}
\end{figure}

\noindent
\textbf{Observations:} 
We now analyze the impact on $Q_{43}$ for the following four windows: (a) \textbf{mem\_live\_2} is the period where both $Q_{43}$ and $Q_{mem-int}$ had begun execution together ($Q_{43}$ had not started in \textbf{mem\_live\_1}), (b) \textbf{mem\_live\_3} shows a window where some other queries had entered the workload, and both $Q_{43}$ and $Q_{mem-int}$ were still running, (c) \textbf{mem\_live\_4} and  \textbf{mem\_live\_5} are the windows when $Q_{43}$ was still running and $Q_{mem-int}$ had finished execution, and (d) \textbf{mem\_live\_total} analyzes the overall impact on $Q_{43}$ from top-$5$ sources during its complete execution window. While $Q_{43}$ was running concurrently with multiple other queries, $Q_{mem-int}$ causes more than 40\% share of the total impact received once it begins as shown in Figure~\ref{fig:indc_mem_dor}. Note that $Q_{43}$ scans \textit{store} and \textit{store\_sales}, whereas we cached \textit{web\_sales} in our induction query. 

\vspace{-2.5mm}
\subsection{Frequency of Data Collection}	 
\begin{figure}[h]
	\vspace{-3mm}
	\centering
	\includegraphics[width=2.7in,keepaspectratio]{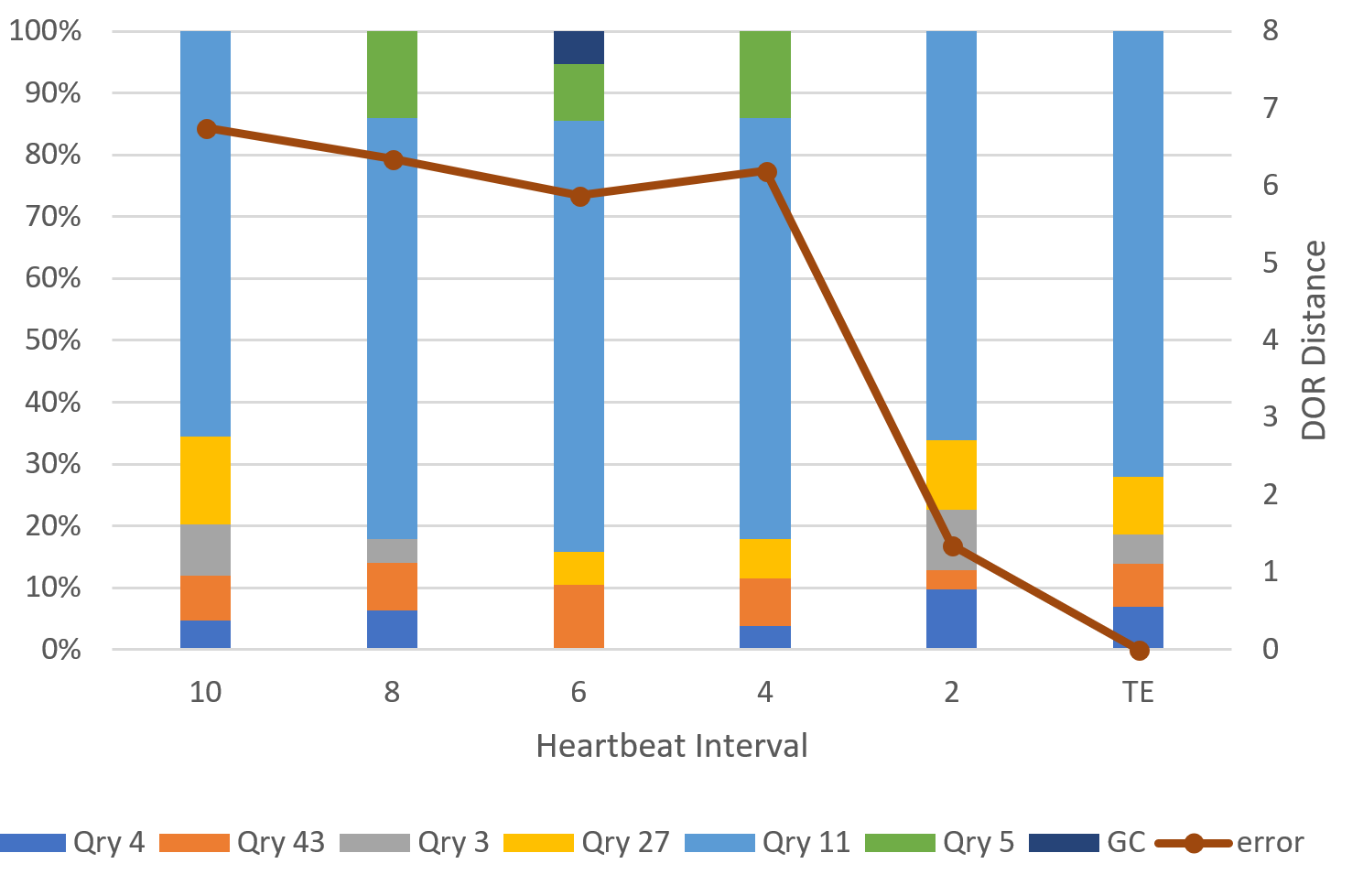}
	\caption{\textbf{\small  Impact of varying intervals of metrics collection on explanation \DOR s. ($TE$ denotes the metrics collection at task-event boundaries (Section~\ref{subsec:freq}) in addition to 2s logging.}}
	\label{fig:VaryingHeartBeatImg}
	\vspace{-7mm}
\end{figure}

\vspace{-2.5mm}
In this experiment, we analyze the impact of varying the frequency of collecting time-series data on the quality of our explanations (\ie \DOR\ values of the sources) towards a \tq 's performance. Figure~\ref{fig:VaryingHeartBeatImg} shows that as the intervals reduce from 10s-2s, the accuracy of the \DOR\ values for all queries (calculated using vector distance) improves and approaches to our ideal case of $TE$ (task-event and 2s interval collection). This value is highly sensitive to the task median runtimes of the workload. For example, a 8s heartbeat time-series collection can miss the impacts on 6s tasks. Our task-event metrics collection approach, thus, enables \proto\ to generate more accurate blames. We next discuss the overheads associated with this instrumentation.  

\begin{figure*}[t]
	\vspace{-2.5mm}
	\centering
	\begin{subfigure}{.35\textwidth}
		\includegraphics[width=2.2in,keepaspectratio]{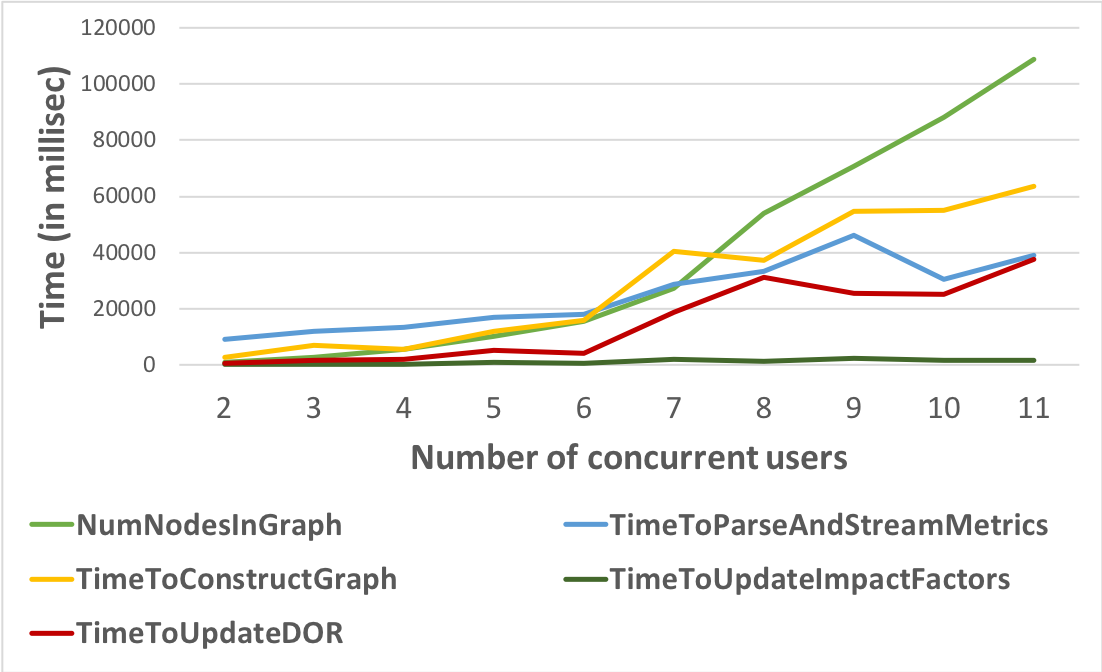}
		\caption{}
		\label{fig:scalability_graph}
	\end{subfigure}%
	\begin{subfigure}{.35\textwidth}
		\includegraphics[width=2.2in,keepaspectratio]{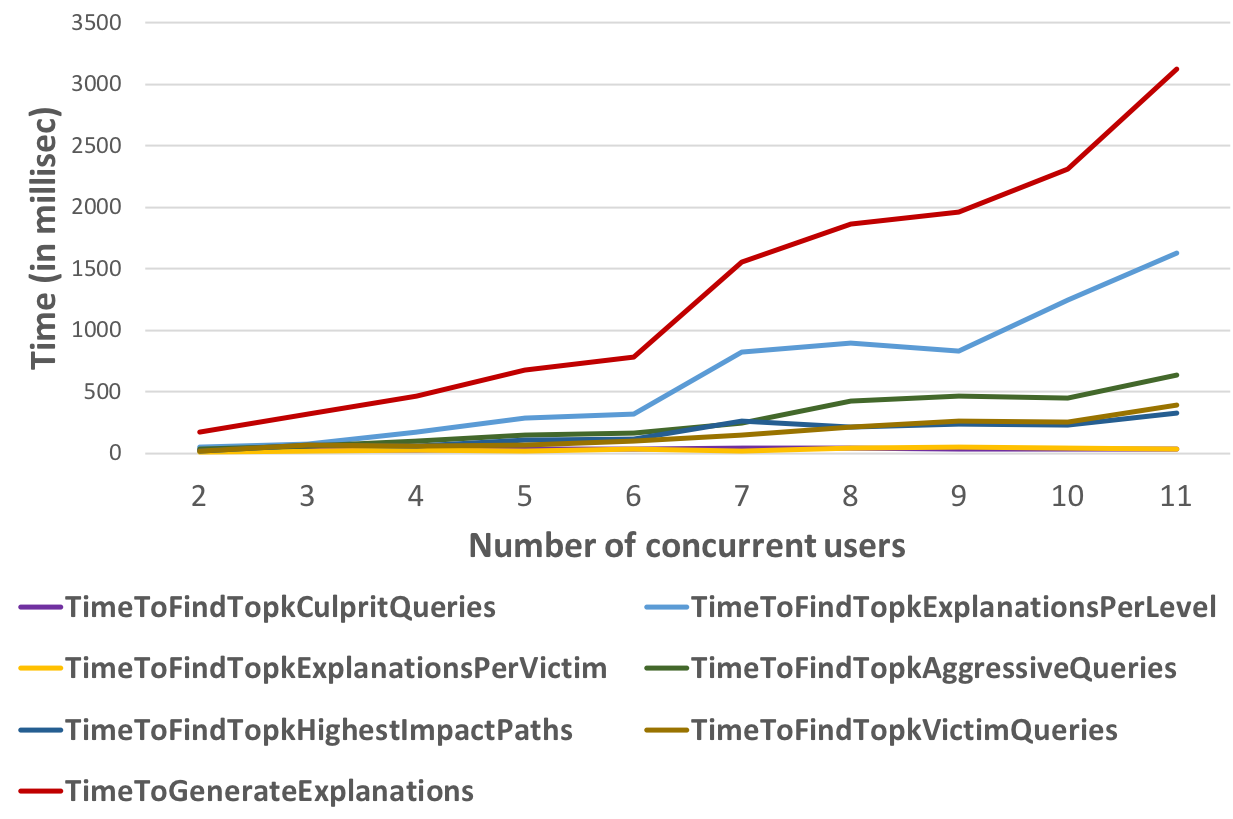}
		\caption{}
		\label{fig:scalability_explanations}
	\end{subfigure}%
	\begin{subfigure}{.35\textwidth}
		\includegraphics[width=2.1in,keepaspectratio]{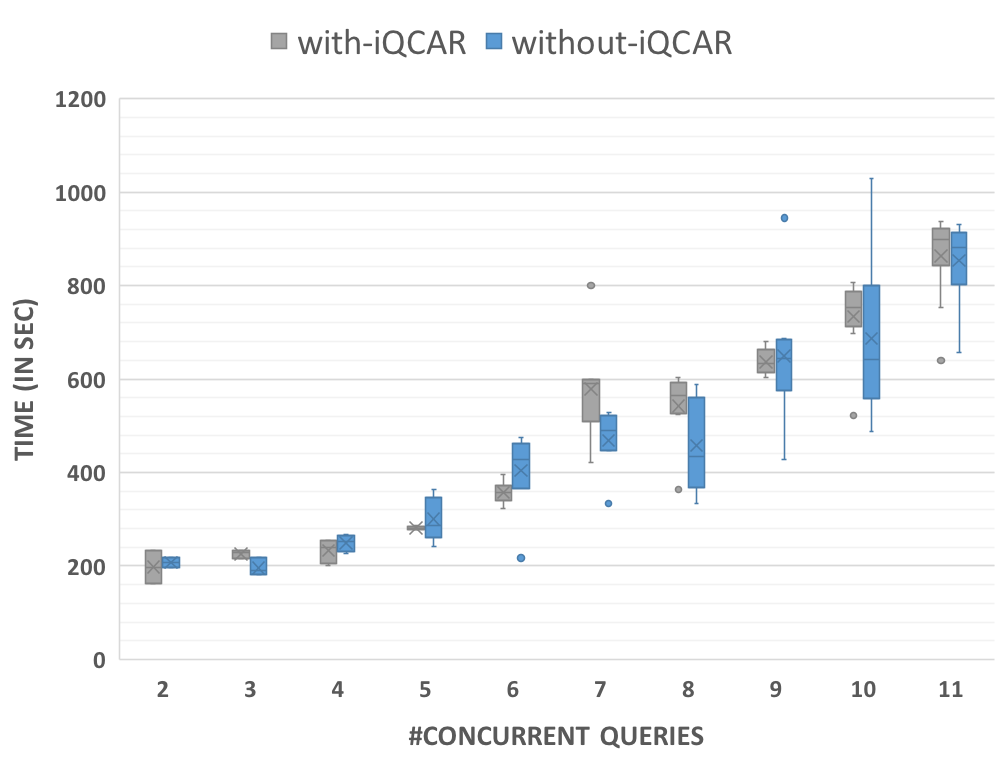}
		\caption{}
		\label{fig:inst_runtime}
	\end{subfigure}%
		\vspace{-4.5mm}
	\caption{\textbf{\small Time taken by the Graph Constructor (Part (a)) and the Explanations Generator (part (b)) modules in various steps. Part (c) shows the overhead of our instrumentation on the runtime of queries in the workload. }}
	\label{fig:instrumentation}
	\vspace{-6.5mm}
\end{figure*}

\vspace{-3mm}
\subsection{Instrumentation}
\vspace{-2mm}
Figure~\ref{fig:inst_runtime} compares the runtime of our workload in Spark with (grey) and without (blue) our instrumentation for publishing time-series for the metrics listed in Section~\ref{subsec:resources}. The heartbeat was set to 2s and task-event metrics collection was set to true - a scenario of maximum load that our instrumentation can put. \par

Since every node in Levels 2, 3 and 4 corresponds to a particular \ts, the subgraph formed by these nodes for one \ts\ at Level 1 is disjoint from the subgraph formed by the nodes for another \ts. They are connected back at Level 5 if multiple \ts s execute  concurrently with the same \sos. This property enables us to construct the subgraphs from Level 0 to Level 4 in parallel\cut{ for each \tq\ to be analyzed}. Figure~\ref{fig:scalability_graph} shows the time taken in parsing time-series data and creating our data model, constructing graph, and times to update the edge weights (\IF) and node weights (\DOR). Figure~\ref{fig:scalability_explanations} presents the times spent for different types of analysis algorithms. From our experience, the runtime of this \textit{Graph Constructor} module dominates the overall latency of outputting explanations in \proto. \par

\begin{table}[ht]
	\vspace{-2mm}
	\small
	\centering
	\begin{tabular}{|l|l|l|l|l|l|l|} \hline
		\multirow{2}{*}{User-Study} &
		\multicolumn{2}{c}{Novice} &
		\multicolumn{2}{c}{Skilled} &
		\multicolumn{2}{c|}{Expert}  \\ \cline{2-7}
		& M & P & M & P & M & P  \\ \hline
		Easy & 28 & 80 & 0 & 100 & 0 & 75  \\ \hline
		Medium & 14 & 40 & 0 & 66 & 16  & 60 \\ \hline
		Hard & 16 & 70 & 0 & 66 & 33 & 66 \\ \hline 
	\end{tabular}
	\caption{Percent correct answers - Manual (M) vs \proto\ (P)}
	\label{table:user-study}
	\vspace{-5mm}
\end{table}

\vspace{-4mm}
\subsection{User Study}
\label{subsec:user-study}
\vspace{-2.5mm}
\cut{To evaluate the expressiveness of \proto\ in presenting explanations for resource interferences to the user,} We performed a user study involving participation from users with varying levels of expertise in databases, Spark and in using system monitoring tools like Ganglia. Briefly, we categorize our users based on their combined experience in all three of these categories as (a) \textit{Novice} (e.g. undergrad or masters database student or users with $\le$ 1 yr experience), (b) \textit{Skilled} (e.g. industry professionals with 1-3 yrs experience) and (c) \textit{Expert} (e.g. DB researcher or industry professionals with more than 3 yrs experience). As such, we had 4 \textit{Experts}, 2 \textit{Skilled} and 4 \textit{Novice} users. This web-based tool enabled the users to analyze a pre-executed micro-benchmark to answer an online questionnaire consisting of 10 multiple-choice questions with mixed difficulty levels. We categorized our questions as \textit{Easy}, \textit{Medium} and \textit{Hard} based on the number of steps involved and the time it took to answer them manually by one of our \textit{Experts}. At the beginning of study, each user was randomly assigned 5 questions to be answered with manual analysis using existing monitoring tools \cite{SparkUI,Ganglia}. The users were then shown the \proto\ user-interface (UI) to answer the rest of the 5 questions. Each question was presented progressively and the time taken to answer it was recorded in order to distinguish between genuine and casual participants. Note that in the \proto\ UI, we purposefully refrained from showing the users the final answers generated by \proto \cut{ for the top-$k$ explanations at each level of the graph}; instead, we presented them with plots based on (a) the measured wait-time and data processed values and (b) consolidated \DOR\ values at each level that aided the users in selecting their answers using \proto\ tool. This activity enabled us to match the answers generated from \proto\ with those selected by our \textit{Expert} users with manual analysis. \par 

In total, users attempted 54\% of our questions requiring manual analysis since they gave up on the remaining questions, whereas they answered 98\% of the questions using \proto\ interface. They got 37\% answers correct with manual analysis, and 69\% right using \proto. Table ~\ref{table:user-study} summarizes the results based on levels of expertise. It shows that using only visual aids on impact values and resource-to-host \RATP\ heatmaps, the accuracy of selecting correct answers increased for each type of user compared to manual analysis. One interesting observation from our user-study is that once the users were finally given an opportunity to compare and review their chosen answers with those generated by \proto\ supplemented with a reasoning, they agreed to accept the output generated by \proto\ for 90\% of the mis-matched answers. In our results, the time taken by the users in answering the questions manually was about 19\% lesser than the time they took to answer using \proto\ UI. When we explored further to understand this counter-intuitive result, we noticed that irrespective of user expertise, most of them spent significant time in answering the first one or two questions but then gave up on the answers too easily for later questions. Whereas, they spent time in exploring the \proto\ UI to attempt more answers.  

\cut{For each of the questions color-coded according to their difficulty levels, Figure ~\ref{fig:user-study-timing} shows the improvement in analysis time using \proto\ is encouraging compared to the time users took when answering the same questions with existing tools. It further shows that even \textit{Expert} users spent much lesser time analyzing with \proto\ than with existing tools.}

\vspace{-5mm}
\section{Conclusion}
\label{sec:conclusion}
\vspace{-2.5mm}
Resource Interferences due to concurrent executions are one of the primary and yet highly mis-diagnosed causes of query slowdowns in shared clusters today. This paper discusses some challenges in detecting accurate causes of contentions, and illustrates with examples why blame attribution using existing methodologies can be inaccurate. To this effect, we propose a theory for quantifying blame for slowdown, and present techniques to filter genuine concurrency related slowdowns from other known and unknown issues. We further showed how our graph-based framework allows for consolidation of blame and generate explanations allowing an \admin\ to explore the contentions and contributors of these contentions systematically. \par

\cut{In this paper, we proposed a model that uses distributions of wait-time per unit data processed by tasks as a basis for formalizing blame attribution to concurrent queries. We further showed how our graph-based framework allows for consolidation of blame across multiple levels of explanations allowing an \admin\ to explore the contentions and contributors of these contentions systematically. Finally, we illustrated how the top-$k$ explanations and rules output by our blame analysis API enables them to find anomalies in query schedules or disproportionalities in resource allocations. \par
 
\noindent 
\textbf{Discussion:} \cut{Research in the area of diagnosing root causes for performance degradation is well established~\cite{Perfxplain,DBSherlock,Driad}. Use of \RATP\ alone lacks the ability to account for wait-time values due to any external known causes or systemic issues. However, the focus of this paper is in establishing a foundation for blame analysis for dataflow-based workload execution which can help the cluster administrators in identifying the scope for rectification in query schedules. }While our current implementation is based on offline analysis after queries have completed execution, we show that this descriptive model still opens a huge space for contention exploration and blame attribution. An \admin\ for a cluster scheduler can use the automated tips suggested by our extensible rule generator in managing her workload better. \cut{Our framework allows valuable consolidation of wait-time data and impact metrics at multiple levels and, thus, opens up many avenues for research in finding schedule anomalies by extending our rules generator.} This feature is particularly useful for well-known or recurring executions. Our on-going research in designing an online contention analysis system, \proto -Live, aims to address the limitations discussed in this paper: that is, sensitivity of \RATP\ to the availability of wait-time data, and application of rules automatically upon sensing interferences. \cut{It involves devising sampling strategies that can still help keep the cost of instrumentation low with high accuracy for \RATP\ values, and automates the application of rules upon sensing interferences.} \par
}
\cut{Finally, \RATP\ requires accurate collection of wait time distributions and data processed values. This adds to the cost of instrumentation since capturing the wait-times of concurrent tasks at overlapping points is of $\mathcal{O}(n^{2})$ complexity. However, we show through our experiments that an assumption of uniform wait-time distributions is reasonable. }	

\vspace{-1.5mm}

\cut{
 We intend to extend 
 this work on offline contention analysis to \proto-\textbf{\textit{Live}}, which will provide online performance insights while queries are running. 
Our on-going research also aims to develop a model for automating dynamic priority levels and delays in stage submissions to devise a contention-aware resource allocation scheme. Another future direction 
 is to investigate impact due to sharing of stages between multiple queries (skipped stages in Spark). 
}

  

\newpage


\end{document}